\documentclass{aa}       
\usepackage{graphicx}
\usepackage{amssymb}
\usepackage{amsmath}
\usepackage{txfonts}

\def\infint{\int_{-\infty}^\infty}
\def\ft{\mathcal{F}}

\begin{document}

\title{Resolution limits in astronomical images}

\author{A.P. Lobanov}

\institute{Max-Planck-Institut f\"ur Radioastronomie,
           Auf dem H\"ugel 69, Bonn 53121, Germany}

\titlerunning{Resolution limits in astronomical images}
\authorrunning{Lobanov}

\offprints{A.P. Lobanov, \email{alobanov@mpifr-bonn.mpg.de}}

\date{Received <date> / Accepted <date>}

\abstract{A method is introduced to derive resolution criteria for
various {\em a priori} defined templates of brightness distribution fitted
to represent structures and objects in astronomical images. The
method is used for deriving criteria for the minimum and maximum
resolvable sizes of a template.  The minimum resolvable size of a
template is determined by the ratio of $(SNR-1)/SNR$, and the maximum
detectable size is determined by the ratio of $1/SNR$. Application of
these criteria is discussed in connection to data from
filled-aperture instruments and interferometers, accounting for
different aperture shapes and the effects of Fourier sampling,
tapering, apodization and visibility weighting. Specific resolution
limits are derived for four different templates of brightness
distribution: (1)~two-dimensional Gaussian, (2)~optically thin
spherical shell, (3)~disk of uniform brightness, and (4)~ring.  The
limiting resolution for these templates changes with SNR similarly to the
quantum limit on resolution.  Practical application of the resolution
limits is discussed in two examples dealing with measurements of
maximum brightness temperature in compact relativistic jets and
assessments of morphology of young supernova remnants.
\keywords{methods: data analysis -- methods: analytical --
techniques: high angular resolution -- techniques: interferometric}}

\maketitle

\section{Introduction}

Accurate knowledge of resolution limits is one of the important
aspects of astronomical observations and data analysis.  There is a
number of research fields in which a quantitative assessment of
resolution limits is vital for interpretation of observations. Most
celebrated examples include the star/galaxy separation in imaging
surveys (e.g. Stoughton et al. \cite{sto2002}), measurements of
stellar diameters (e.g. Tango \& Davis \cite{td2002}), measurements of
sizes and shapes of young supernova remnants (e.g. Marcaide et
al. \cite{mar+95}, McDonald et al. \cite{mcd+2001}, P\'erez-Torres et
al. \cite{per+2002}, Bartel \& Bietenholz \cite{bb2003}), and
estimates of brightness temperatures in extragalactic jets based on
very long baseline interferometry (VLBI) measurements of sizes of most
compact regions (cores) of the jets (e.g.  Moellenbrock et
al. \cite{mol+1996}, Lobanov et al. \cite{lob+2000}.  Horiuchi et
al. \cite{hor+2004})

The Rayleigh (\cite{ray1879}) resolution criterion and its several
modifications are commonly considered as a measure of limiting
resolution. The Rayleigh limit is expressed by the resolution distance
$R = 1.22\,(\lambda/2\alpha_\mathrm{f})$, where $\lambda$ is the wavelength of
light and $\alpha_\mathrm{f} = D/f$ is the ratio of the diffracting aperture,
$D$, to the distance, $f$, between the aperture and the image
plane. Strict application of the Rayleigh criterion is limited to the
specific case of resolving two point-like objects of comparable
brightness. In terms of Fourier optics, resolution is better
represented by the {\em instrumental bandwidth} that represents the
range of Fourier frequencies sampled by an imaging apparatus.
Diffraction results in existence of the maximum (cutoff) frequency
$q_\mathrm{max}$ that limits the resolution. Indeed, the Rayleigh
limit has been reformulated, in Fourier optics, in terms of the {\em
Nyquist distance} $R = \pi/q_\mathrm{max}$.  Linear inversion methods
applied to {\em out-of-band extrapolation} yield resolution limits
smaller than the one prescribed by $R$. This provides the basis of the
so-called ``superresolution'' techniques (Bertero \& De Mol
\cite{bd1996} and references therein).

 In many practical cases, an observed brightness distribution has to
be tested against a specific, extended pattern, for instance, a
characteristic surface brightness distribution of a galaxy, stellar
disk or supernova shell.  A number of astronomical studies employ
fitting image brightness distribution by such pre-defined templates,
with the definitions being based on physical models or previous
observational evidence. Introducing {\em a priori} knowledge (or a
postulate) of the brightness distribution of an emitting region is
similar to using the out-of-band extrapolation (in which a definitve
postulate is employed about the shape of the emitting region).  In
both cases, the limiting resolution depends on the signal-to-noise
(SNR) of detection. In contrast to the out-of-band extrapolation, the
limiting resolution for {\em a priori} defined templates is determined
ultimately by quantum fluctuations of light (Kolobov \& Fabre
\cite{kf2000}), because the templates are represented by analytic
functions that can be extended over an infinite range of frequencies
in Fourier domain.

An analytical criterion for calculating resolution limits can be
derived for specific templates of brightness distribution (TBD,
hereafter) from their respective Fourier visibility representations
(see e.g. Thompson, Moran, Swensson \cite{tms86}, Taylor, Carilli,
Perley \cite{tcp99}). Working in the Fourier plane has a
specific advantage of separating positional information (contained in
the visibility phase) from information about the extent of the
emitting region (contained in visibility amplitudes). This has been
recognized early on in radio interferometry and used for a variety of
purposes, most notably for error estimates (see e.g. Fomalont
\cite{fom91}).

Analysis of visibility distribution in Fourier domain is applied in
this paper to derive general resolution criteria and obtaining
resolution limits for specific TBD.  The basic foundations of the
method for deriving the resolution limits are summarized in
Sect.~\ref{sc:basic}. The general resolution criteria are introduced in
Sect.~\ref{sc:criterion} for determining the minimum and maximum
resolvable size of a TBD. Several relevant TBD are described in
Sect.~\ref{sc:templates}. Their respective limiting resolutions are
calculated in Sect.~\ref{sc:limits} and compared with the quantum
limits on resolution. Applications of the resolution limits are
outlined, in Sect.~\ref{sc:applications}, with two examples dealing
with studies of brightness temperature in compact AGN and assessment
of morphology of young supernovae.

\section{Basic definitions}
\label{sc:basic}

Consider an emitting region of finite size that has an integral flux
density $F$ and a brightness distribution $I(l,m)$.  The brightness
distribution $I(l,m)$ is measured in the image plane, which is
described by a rectangular coordinate system $(l,m)$. The
corresponding Fourier (or interferometric) visibility distribution
$V(u,v)$ is measured in Fourier plane ($uv$-plane), with the
respective rectangular coordinate system $(u,v)$. The two are related
via Fourier transform ($\ft$):
\[  
V(u,v) = \ft\, I(l,m) = \infint \infint I(l,m) \exp[-2\pi\, i(u\,l+v\,m)] \, 
\mathrm{d}x \, \mathrm{d}y\,,
\]
\[
I(l,m) = \ft^{-1} V(u,v) = 
\infint \infint V(u,v) \exp[2\pi\, i(u\,l+v\,m)] \, \mathrm{d}x \, \mathrm{d}y\,.
\]
The visibility
distribution is complex-valued, with the visibility amplitude determining
the limiting resolution and the visibility phase affecting ultimately the 
positional accuracy in the image plane.
For a circularly symmetric brightness distribution, the relation between
$V(u,v)$ and $I(l,m)$ can be represented by Hankel transforms 
\[
V(q) =
\int_0^{\infty} I(r) J_0(2\pi q\, r)r\, \mathrm{d} r\,, 
\]
\[
I(r) =
\int_0^{\infty} V(q) J_0(-2\pi q\, r)q\, \mathrm{d} q\,, 
\]
where $J_{0}$ is the zero order Bessel function and $q = (u^2 +
v^2)^{1/2}$ and $r = (x^2 + y^2)^{1/2}$ are radial coordinates in the
Fourier plane and image plane, respectively.  For simplicity of
derivation, circular symmetry will be assumed throughout
this paper, but the method of analysis is valid for the general case as well,
which can be readily shown by expanding the variables $r$ and $q$ into
their two dimensional representations $(l,m)$ and $(u,v)$.

The relevant range ($q_\mathrm{min},q_\mathrm{max}$) of frequencies
(coordinate dimensions) in the visibility plane is specific to each
individual observational setup. In most of the practical cases,
$q_\mathrm{min}\rightarrow 0$ or $q_\mathrm{min} \ll q_\mathrm{max}$
can be assumed.  For filled-aperture instruments, $q_\mathrm{max}$ is
determined by the ratio $D/\lambda_\mathrm{obs}$ of the diameter of
the aperture to the observing wavelength. For interferometers, the
diameter is substituted by the maximum baseline length,
$B_\mathrm{max}$, between individual elements.  If $q_\mathrm{max}$ is
known, the Fourier transform of the brightness distribution in an
astronomical image can be used for deriving rigorous resolution
criteria for a TBD. To convert the result to angular units in the
image plane, $q_\mathrm{max}$ has to be related to the size of the
point spread function (PSF) or the synthesised beam, in the
terminology adopted in radio interferometry.  This can be achieved by
considering the interferometric visibility sampling function (SF) ---
an effective coverage of the Fourier plane, which is often also termed
as the transfer function. The theoretical shape of the PSF is
completely determined by the combination of the SF and the frequency
range ($q_\mathrm{min},q_\mathrm{max}$). In practice, the shape and
size of the PSF are further modified by aberrations, weighting and
tapering (apodization) of data (e.g. Norton \& Beer \cite{nb76},
Briggs \cite{bri95}, Briggs, Schwab \& Sramek \cite{bss99}), which
also affects actual resolution limits. These effects have to be taken
into account, thus reformulating $q_\mathrm{max}$ in terms of the PSF.

\subsection{Sampling of the Fourier domain}

The shape of the PSF, $B(l,m)$, depends on the SF (or {\it
uv}--distribution) $s(u,v)$ in the Fourier domain. The PSF is then, in
the simplest case, $B(l,m) = {\cal F} s(u,v)$.  The PSF is
characterized by its width at half power level (FWHM), which is often
also termed the ``half power beam width''(HPBW, hereafter).
Generally, the two-dimensional shape of the PSF must be considered,
and resulting directional resolution limits should be derived.  If
this is not feasible, an equivalent circular HPBW, ${\cal
B}_\mathrm{c}$, can be used, which depends on the actual shape of the
PSF or, more generally, on the shape of the SF. ${\cal B}_\mathrm{c}$
gives then a corresponding size of a circular PSF that contains the
same integrated flux density as the elliptical (or generally speaking,
non-circular PSF and SF).  For an elliptical SF,
\begin{equation}
{\cal B}_\mathrm{c} = (a\, b)^{1/2}\,,
\label{eq:hpbw-e}
\end{equation}
where $a$ and $b$ are the major and minor axes of the PSF. For a
rectangular SF,
\begin{equation}
{\cal B}_\mathrm{c} = (\pi\, a\, b)^{1/2}\,.
\label{eq:hpbw-r}
\end{equation}
Equations~(\ref{eq:hpbw-e}) and (\ref{eq:hpbw-r}) can be applied to
describe ${\cal B}_\mathrm{c}$ of most of the imaging instruments used
in astronomy.

The relation between the beam and the sampling function is even more
complex if $s(u,v)$ provides either incomplete or non-uniform coverage
of the Fourier plane. The visibility sampling can then be
characterized by the smallest and largest {\it uv}--distances
($q_\mathrm{min}$, $q_\mathrm{max}$) and by the density
$\rho_{\sigma}(u,v)$ of the {\it uv}--samples (assuming
$q_\mathrm{max} \gg q_\mathrm{min}$).  The density is
defined so that $\rho_{\sigma}(u,v){\rm d}u\,\mathrm{d}v$ gives the
number of visibilities within the area given by
($u\pm0.5\mathrm{d}u\,\,,v\pm0.5\mathrm{d}v$).  The visibility
sampling can be modified by a weighting function $w(u,v)$ to emphasize
certain fractions of the visibility distribution and optimize image
sensitivity on selected spatial scales.
With the weighting function applied,
\begin{equation}
s(u,v) = \rho_{\sigma}(u,v) w(u,v)\,,
\label{eq:suv}
\end{equation}
giving the total number of visibility data points $N(u,v) = \int\int
s(u,v)\mathrm{d}u\,\mathrm{d}v$.  

Depending on the form of $s(u,v)$, an appropriate correction factor
${\cal S}_\mathrm{q}$ must be applied to the equivalent (or directional)
HPBW ${\cal B}_\mathrm{c}$, which gives
\begin{equation}
q_\mathrm{max} = ({\cal S}_\mathrm{q} {\cal B}_\mathrm{c})^{-1}\,.
\label{eq:qmax-hpbw-cor}
\end{equation}

\subsection{${\cal S}_\mathrm{q}$ for a power law $s(u,v)$}
\label{sc:Qs-powerlaw}

A power-law SF can be used to describe an ideal interferometer as well
as a major fraction of single aperture instruments. The effect of a
power-law SF on the HPBW can be calculated analytically.  Assuming
circular symmetry yields $s(q) = \rho_{\sigma}(q)\, w(q)$.  The
density of samples is $\rho_{\sigma}(q) \propto q^{-2}$.  The function
$w(q)$ is then given by $q^{\beta}$, where $\beta$ is a power index
decribing the weighting.  The choice of $w(q)$ that yields $s(q) =
const$ is commonly known as ``natural weighting''. Thus, $\beta = 2$,
for the natural weighting, and the corresponding $s(q)$ is given by
$\Pi (q)$, where $\Pi(q) = H(1+q) H(1-q)$ and $H$ is the unit
Heaviside function. The natural weighting increases sensitivity to
extended emission at the expense of reducing the resolution of the
image. The resolution is optimized by choosing $w(q) = const$ (uniform
weighting), for which the respective $s(q) = \rho_{\sigma}(q)$.  This
yields $s(q) \propto q^{-2}$ for the uniformly weighted synthesised
interferometric aperture (similar to the result obtained for filled
aperture telescopes).  A general form of the sampling function is then
\begin{equation} 
s(q)= \Pi(q)\,w(q) \rho_{\sigma} q \mathrm{d}q =
\Pi(q)\, q^{\beta-2} \mathrm{d}q\,,
\label{eq:sr}
\end{equation}
with $\beta = 0$ for the uniform weighting and $\beta = 2$ for the
natural weighting. Thus, for a power law $s(q)$ given by Eq.~(\ref{eq:sr}),
\begin{equation}
{\cal S}_\mathrm{q} = 2^{1-\beta/2}\,,
\label{eq:sq-plaw}
\end{equation}
and $q_\mathrm{max}$ is related to the effective HPBW as follows:
\begin{equation} 
q_\mathrm{max} =
\frac{1}{2^{1-\beta/2} {\cal B}_\mathrm{c}}\,,
\label{eq:u-beam}
\end{equation}
where $a$ and $b$ are the major and minor axes of the synthesised beam.
For filled aperture instruments, $\beta = 0$ and ${\cal S}_\mathrm{q}
= 1/2$ can be generally adopted.  $S_\mathrm{q}$ can be modified by
apodization, depending on the shape and parameters of the apodizing
function. The range of $\beta\,\,[0,2]$ corresponds to a range $[-1,1]$
of the {\em robustness parameter} introduced by Briggs (\cite{bri95}).

In interferometers, the shape of the SF may deviate from the power-law
due to several factors including sparsity of receiving elements,
non-optimal configurations, and reduced mutual visibility on very long
baselines.  A relative error $\sigma_\mathrm{q}$ of ${\cal
S}_\mathrm{q}$ due to an incomplete SF of an interferometer can be
approximated from the size of gaps $\Delta q/q$ in the SF. For an
interferometer with $N$ regularly spaced elements, $\sigma_\mathrm{q}
\approx (\Delta q/q)^2 = [(q_\mathrm{max}/q_\mathrm{min})^{3/N} -
1]^2$ (Lobanov
\cite{lob03}\footnote{http://www.skatelescope.org/PDF/ska\_memo38.pdf}).
The resulting $\sigma_\mathrm{q}$ is a few per cent or less for most of
the existing connected interferometers. It is expected to be improved
substantially in the future interferometric instruments such as ALMA
and SKA. In present-day VLBI observations, $\sigma_\mathrm{q}$ remains
about 10\%--15\%. Thus, Eq.~(\ref{eq:sq-plaw}) provides a relatively
accurate assessment of ${\cal S}_\mathrm{q}$. Numerical analysis of the
specific instrumental SF should be employed whenever a better accuracy
is required.

\subsection{Noise and SNR}

The Fourier transform has a fundamental property of preserving the
covariance of noise. It implies that the signal-to-noise ratio (SNR)
measured in the image plane is the same as the one that would have
been recovered if the same structure was fitted to the data in the
visibility plane.  Then $SNR = F/\sigma_\mathrm{n} =
V(0)/\sigma_\mathrm{n}$, which is the ratio of the integral flux
density to the variance of the additive white Gaussian noise with a
power spectrum $\rho(q) = \sigma^2_\mathrm{n}\, \delta(q)$.  This
property enables the derivation of resolution criteria from the
Fourier transform of the image brightness distribution. It should be
noted that the definition of SNR introduced above is different from
another commonly used definition given by $I(0)/\sigma_\mathrm{n}$.

The full SNR is applied for calculation of the resolution limits if
the brightness distribution is fit by a single template or if the
position of the template of interest is fixed, in a fit by multiple
templates.  If the brightness distribution is fit by multiple
templates at variable positions, the SNR in the estimates of
$d_\mathrm{lim}$ and $d_\mathrm{res}$ should be multiplied by a factor
$\eta = \sqrt{1 - \sigma_\mathrm{phase}^2/ \sigma_\mathrm{amp}^2}$,
where $\sigma_\mathrm{phase}$ and $\sigma_\mathrm{amp}$ is the noise in
the visibility phases and amplitudes, respectively. Typically, the
noise is distributed evenly between the amplitudes and phases, and so
$\sigma_\mathrm{phase} = \sigma_\mathrm{amp} =
\sigma_\mathrm{n}/\sqrt{2}$ and $\eta = 1/\sqrt{2}$.

\section{Resolution criteria}
\label{sc:criterion}

Consider a circularly symmetric object in the image plane, which has
 an integral flux density $F$ and a distribution
 of brightness $I(r)$ that transforms into a visibility distribution
 $V(q)$. The noise is given by $\sigma_\mathrm{n}$. The
 visibility distribution $V(q)$ is normalized by $F$, implying
 $V(0) \equiv 1$.   The visibility
 distribution can be conveniently represented as
\begin{equation}
V(q) = d\, {\cal V}_\mathrm{s}\,\,q\,,
\label{eq:vq-terms}
\end{equation}
where ${\cal V}_\mathrm{s}$ and $d$ represent the shape and
size of a TBD, respectively. 

\subsection{Minimum resolvable size}
\label{sc:dlim}

For a point-like brightness distribution, $I(r) = \delta(r)$,
$V(q)\equiv 1$ at all $q$. For an object with extended brightness
distribution, $V(q) = 1 - \xi(q)$, where $\xi(q)$ depends on the size
and shape of the object. If this deviation of the visibility
distribution from unity can be detected in the data, then the
brightness distribution $I(r)$ is resolved. The function $\xi(q)$
evaluated at $q_\mathrm{max}$ can be related to the SNR of detection
of the TBD. The resolution criterion is then given by
$\xi(q_\mathrm{max}) \ge \sigma_\mathrm{n} = 1/SNR$.
This criterion implies that $I(r)$ is resolved if
\begin{equation}
V(q_\mathrm{max}) \le \frac{SNR-1}{SNR} \equiv f_\mathrm{m}(SNR)\,.
\label{eq:res-cond}
\end{equation}
Solving this equation for $d$, recovered from the expression for
$V(q_\mathrm{max})$, yields the limiting resolution $d_\mathrm{lim}$
for a given template. Following Eq.~(\ref{eq:vq-terms}), $V(q_\mathrm{max})
= d_\mathrm{lim}\,q_\mathrm{max}\,{\cal V}_\mathrm{s}$, and the
limiting resolution is
\begin{equation}
d_\mathrm{lim}  = \frac{1}{q_\mathrm{max}}\frac{f_\mathrm{m}(SNR)}{{\cal V}_\mathrm{s}}\,.
\label{eq:dlim-gen}
\end{equation}
The form ${\cal V}_\mathrm{s}$ depends on the choice of the template.  
The ratio 
\begin{equation}
{\cal R}_\mathrm{lim} = \frac{f_\mathrm{m}(SNR)}{{\cal V}_\mathrm{s}}
\label{eq:rlim}
\end{equation}
provides a theoretical resolution factor independent from the size and
shape of the PSF of the data.  ${\cal R}_\mathrm{lim}$ depends only on
the SNR and the shape of a TBD.  ${\cal R}_\mathrm{lim}$ is
a measure of the space-bandwidth product $c =
d\,q_\mathrm{max}$. Indeed ${\cal R}_\mathrm{lim} =
2\,c_\mathrm{lim}$.  This elucidates the connection between ${\cal
R}_\mathrm{lim}$ and the Shannon number, $S=d/R$, which gives, for a
spatial interval $[-d/2,\,+d/2]$, the number of sampling points that
are separated by the Nyquist distance $R=\pi/q_\mathrm{max}$ (Toraldo
di Francia \cite{tor1969}). It follows immediately that
\begin{equation}
{\cal R_\mathrm{lim}} = \pi\,S_\mathrm{lim}\,,
\label{eq:dlim-shannon}
\end{equation}
where the subscript in $S$ implies that the Shannon number is evaluated at $d_\mathrm{lim}$.
Recalling Eq.~(\ref{eq:qmax-hpbw-cor}), the resolution limit can be
rewritten as
\begin{equation}
d_\mathrm{lim} = {\cal R}_\mathrm{lim}\,{\cal S}_\mathrm{q} \,{\cal B}_\mathrm{c}\,.
\label{eq:dlim-fin}
\end{equation}
The
HPBW given by ${\cal B}_\mathrm{c}$ can represent both the FWHM
of a non-circular PSF along a selected direction in the image plane or an
equivalent of the FWHM of a circular PSF calculated using
Eqs.~(\ref{eq:hpbw-e}--\ref{eq:hpbw-r}). The term
${\cal S}_\mathrm{q}$ accounts for the effect of limited sampling of
the Fourier domain (see Sect.~\ref{sc:Qs-powerlaw}).

\subsection{Largest detectable size}
\label{sc:dres}

Visibility amplitude of a completely resolved template reaches
$\sigma_\mathrm{n}$ at a scale $q_\mathrm{res}$ such that
$q_\mathrm{res}\le q_\mathrm{max}$. This yields a resolution
condition
\begin{equation}
V(q_\mathrm{res}) = \sigma_\mathrm{n} = \frac{1}{SNR} \equiv f_\mathrm{r}(SNR)\,.
\label{eq:maxres}
\end{equation}
The largest detectable size is obtained by using Eq.~(\ref{eq:vq-terms}) to
separate different terms in $V(q_\mathrm{res})$, which gives
\begin{equation}
d_\mathrm{res} = \frac{1}{q_\mathrm{res}}\frac{f_\mathrm{r}(SNR)}{{\cal V}_\mathrm{s}}\,,
\label{eq:dmax}
\end{equation}
similar to Eq.~(\ref{eq:dlim-gen}). The main difference from
Eq.~(\ref{eq:dlim-gen}) is that $f_\mathrm{r}(SNR)/{\cal
V}_\mathrm{s}$ cannot now be used as an SNR-dependent resolution
factor separately from $q_\mathrm{res}$, since the latter also depends
on the SNR of the detection. Evidently, $\mathrm{SNR} =
\sqrt{N(q_\mathrm{max})/N(q_\mathrm{res})}$ is required for a fully
resolved template to be detected at $q_\mathrm{res}$. Here,
$N(q_\mathrm{max})$ and $N(q_\mathrm{res})$ are the numbers of
visibility samples within $q_\mathrm{max}$ and $q_\mathrm{res}$,
respectively. The value of $q_\mathrm{res}$ depends on the sampling
function $s(q)$, and it can be found by solving the equation $N(q) =
\int_{0}^{q} s(q) \mathrm{d}q$ for the upper limit of integration.

Alternatively, $q_\mathrm{res}$ can be 
expressed as a combination of SNR and $q_\mathrm{max}$, with the latter being
conveniently related to the HPBW using Eq.~(\ref{eq:qmax-hpbw-cor}).
For instance, for a power-law $s(q)$ given by Eq.~(\ref{eq:sr}), 
\begin{equation}
q_\mathrm{res} = \frac{q_\mathrm{max}}{\mathrm{SNR}^{1/(3-\beta)}}\,,
\label{eq:qres-plaw}
\end{equation}
and the largest resolvable size is then given by
\begin{equation}
d_\mathrm{res} = \frac{\mathrm{SNR}^{1/(3-\beta)} f_\mathrm{r}(SNR)}
{{\cal V}_\mathrm{s}}\,{\cal S}_\mathrm{q}\, {\cal B}_\mathrm{c}\,.
\label{eq:dmax-plaw}
\end{equation}
An equivalent of the SNR-dependent resolution factor ${\cal R}$ can
now be defined by
\begin{equation}
{\cal R}_\mathrm{res} = \frac{\mathrm{SNR}^{1/(3-\beta)} f_\mathrm{r}(SNR)}{{\cal V}_\mathrm{s}}\,,
\label{eq:rres}
\end{equation}
transforming Eq.~(\ref{eq:dmax-plaw}) into the same form as
Eq.~(\ref{eq:dlim-fin}). By analogy with Eq.~(\ref{eq:dlim-shannon}),
\begin{equation}
{\cal R}_\mathrm{res} = \pi\,S_\mathrm{res}\,.
\end{equation}

The limit $d_\mathrm{res}$ signifies the largest size of a template
that can be detected. At sizes larger than $d_\mathrm{res}$,
sensitivity to extended emission becomes insufficient to detect the
specific shape of brightness distribution described by the template.
In case of fitting astronomical data by multiple templates (e.g. in
``modelfitting'' used in radio interferometry; see Pearson
\cite{pea99}), $d_\mathrm{res}$ gives an estimate of the size at which
a template must be split into two templates, to maintain
an adequate representation of the structure observed. The ratio
$d_\mathrm{res}/d_\mathrm{lim}$ defines an effective ``resolution
dynamic range'' (RDR), ${\cal D}$, that expresses the range of template sizes
that can be detected at a given SNR. The RDR depends
only on the SNR of detection and the data weighting, since the terms
related to HPBW, SF, and the shape of the template (${\cal
V}_\mathrm{s}$) all cancel in the ratio, giving
\begin{equation}
{\cal D} = \frac{{\cal R}_\mathrm{res}}{{\cal R}_\mathrm{lim}}\,.
\label{eq:rdr}
\end{equation}

\section{Relevant templates of brightness distribution}
\label{sc:templates}

There are several commonly used TBD for describing structures
observed in astronomical images (particularly in VLBI images). These
include a two-dimensional Gaussian profile (``Gaussian component''),
a disk of uniform brightness, an optically thin sphere (or shell of finite
thickness), and a ring (see Pearson \cite{pea99}).  Gaussian components
are commonly used to approximate and quantify structures recovered in
VLBI images of relativistic jets and other objects. Shells, disks and
rings are applied to analysis of images of supernovae and stellar
objects.

\subsection{Gaussian component}

A circular Gaussian shape with the FWHM $d$ is given by 
\begin{equation} 
I(r) = {2\sqrt{\ln 2} \over \sqrt{\pi} d}
\exp\left( {- 4 \ln 2\, r^2\over d^2}\right) = \frac{C_{\sigma}}{d}\exp\left(
  {- \pi C_{\sigma}^2 r^2\over d^2}\right) \,, 
\end{equation} 
with $C_{\sigma} =
2\sqrt{\ln(2)/\pi}$. An elliptical Gaussian component $I(r,\phi)$ can be
described substituting $d$ with the running FWHM 
\begin{equation} 
d_{\phi} = \sqrt { [a \cos(\phi -
  \psi)]^2 + [b \sin(\phi - \psi)]^2}\,, 
\end{equation} 
where $a$, $b$ are the major
and minor axes of the component and $\psi$ is the position angle of the
major axis.   
The visibility distribution corresponding to the Gaussian
component is
\begin{equation} 
V(q)= \exp\left({-(\pi d_{\phi}\,q)^2 \over 4 \ln 2} \right)
= \exp\left(- \frac{ \pi\, d_{\phi}^2 q^2}{C_{\sigma}^2} \right)\,.
\label{eq:Fu}
\end{equation}
An equivalent of circular FWHM can be obtained in this case by finding the
corresponding diameter of a circular Gaussian which has the same area
at the half--power level. This gives $d_\mathrm{FWHM} = \sqrt{a\,b}$.

\subsection{Spherical shell of finite thickness}

A homogeneous shell with the outer radius $r_\mathrm{s}$
and a thickness $\delta_\mathrm{r} r_\mathrm{s} =(1-\alpha) r_\mathrm{s}$ (with
$0\le\alpha\le 1$) is described by 
\begin{equation} 
I(r) =
\begin{cases}
 f_{r_\mathrm{s}}(r) - f_{\alpha r_\mathrm{s}}(r) & r\le r_\mathrm{s}\,,\\
0 & r>r_\mathrm{s}
\end{cases}
\label{eq:shell-r}
\end{equation}
with $f_{z}(r) = \sqrt{z^2 - r^2}$. Thus, $\alpha=0$ describes an infinitely thin shell and $\alpha\rightarrow 1$ describes a filled sphere. The respective visibility distribution is
given by  Hankel transform of Eq.~(\ref{eq:shell-r}), which yields
\begin{equation}
\begin{split}
V(q) =& \xi_\mathrm{V} \frac{1}{8 \pi^3 q^3}\left[ \sin(2\pi\,r_\mathrm{s}\,q) -
2\pi\,r_\mathrm{s}\,q\,\cos(2\pi\,r_\mathrm{s}\,q) - \right. \\
& 
\left.  - \sin(2\pi\,\alpha\,r_\mathrm{s}\,q) + 2\pi\,\alpha\, r_\mathrm{s}\,q \cos(2\pi\,\alpha\,r_\mathrm{s}\,q) \right] \,,\\
\end{split}
\label{eq:shell-u}
\end{equation}
where $\xi_\mathrm{V} = 3/[r_\mathrm{s}^3(1-\alpha^3)]$ is the normalization factor
derived from the condition $V(0)=1$.
If the shell is described by its outer diameter $d= 2 r_\mathrm{s}$, then $f_{z} = 
\sqrt{z^2-d^{2}/4}$, and
\begin{equation}
\begin{split}
V(q) = & \frac{3}{(1-\alpha^3)\pi^3 d^3 q^3}\left[ \sin(\pi\,d\,q) -
\pi\,d\,q \cos(\pi\,d\,q) - \right. \\
& \left.  - \sin(\alpha\,\pi\,d\,q) + \alpha\,\pi\,d\,q \cos(\alpha\,\pi\,d\,q) \right] \,, \\
\end{split}
\label{eq:shell-ud}
\end{equation}

For small $d$ ($d\le{\mathrm{HPBW}}$), $V(q)$
can be approximated by the Taylor expansion
\begin{equation}
V(q)  = 1- \frac{1}{10}\left(\frac{1 - \alpha^5}{1-\alpha^3}
\right) \pi^2 d^2 q^2 + o(q^4)\,.
\label{eq:vis-shell1}
\end{equation}

\subsection{Disk of uniform brightness}

A uniformly bright disk of diameter $d$ is described by
\begin{equation}
I(r) = 
\begin{cases}
4/(\pi\,d^2), & r\le d\,,\\
0 & r> d\,.
\end{cases}
\label{eq:Ir-d}
\end{equation}
The corresponding visibility distribution is 
\begin{equation}
V(q) = \frac{2\,J_{1}(\pi\,d\,q)}{\pi\,d\,q}\,,
\label{eq:Fu-d}
\end{equation}
where $J_{1}$ is the Bessel function of the first kind.  A useful
approximation of $V(q)$ in the small-size limit is given by the Taylor
expansion
\begin{equation}
V(q) = 1 - \frac{(\pi\,d\,q)^2}{8} + o(q)^4\,.
\label{eq:Fu-dexp}
\end{equation}

\subsection{Ring}

In a ring of diameter $d$, the brightness is zero everywhere except on
the circumference:
\begin{equation}
I(r) = \frac{1}{\pi\,d}\delta(r-d/2)\,.
\label{eq:Ir-ri}
\end{equation}
The respective visibility distribution is given by
\begin{equation}
V(q) = \mathrm{J}_0(\pi\,d\,q)\,,
\label{eq:vis-ring}
\end{equation}
and it can be approximated conveniently using the
Taylor expansion of the Bessel function
\begin{equation}
\mathrm{J}_0(\pi\,d\,q) = 1 - \frac{(\pi\,d\,q)^2}{4} + o(q)^4\,.
\label{eq:vis-ring-exp}
\end{equation}

\begin{figure}
\centerline{
\includegraphics[width=0.45\textwidth, bb =28 45 706 522, clip=true]{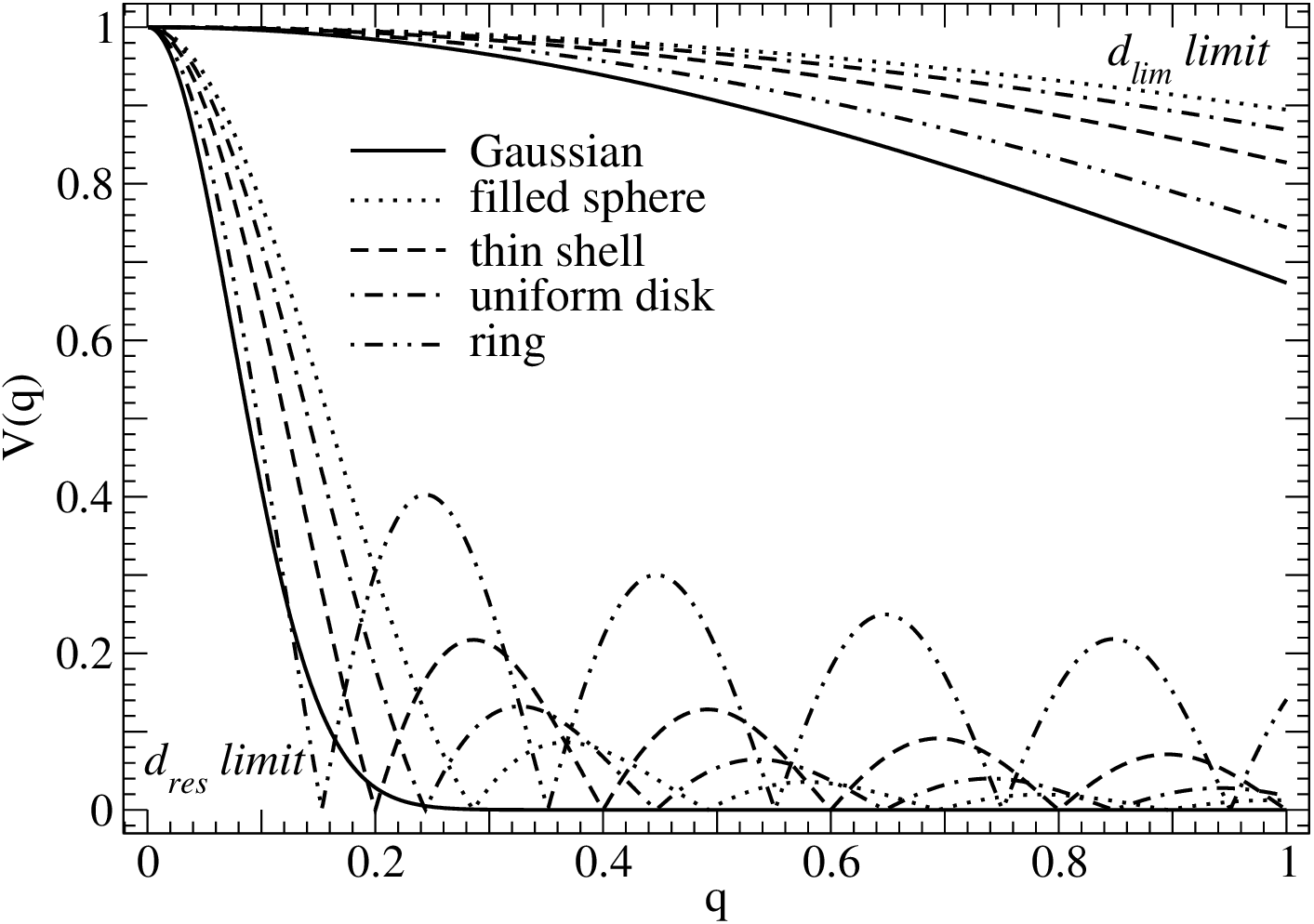}}
\caption{Visibility amplitudes $V(q)$ of the templates plotted for the small-size
($d_\mathrm{lim}$) and large-size ($d_\mathrm{res}$) limits. }
\label{fg:vq}
\end{figure}

The visibility amplitudes of all five templates are compared in
Fig.~\ref{fg:vq}, which illustrates the behavior of $V(q)$ in the
small-size and large-size limits (corresponding to the conditions used for
deriving $d_\mathrm{lim}$ and $d_\mathrm{res}$, respectively. It
should be noted that the third order Taylor expansions used above
approximate well the shape of $V(q)$ only in the small-size
limit. Thus, only $d_\mathrm{lim}$ derived from the Taylor
approximations of $V(q)$ are accurate. In the large-size limit,
analytical expressions obtained with $V(q)$ given by
Eqs.~(\ref{eq:vis-shell1}), (\ref{eq:Fu-dexp}) and
(\ref{eq:vis-ring-exp}) underestimate $d_\mathrm{res}$ by up to
40\%. To remove this inaccuracy, numerical solutions must be obtained
for $d_\mathrm{res}$ of the shell, disk and ring templates. The
numerical solutions can then be used to derive empirical correction
functions for the analytical expressions for $d_\mathrm{res}$. These
correction functions are derived in Appendix~\ref{sc:app-a}. 
The corrected $d_\mathrm{res}$ are accurate to within 0.01\%.

\subsection{Noise term in Taylor expansions of $V(q)$}

The $o(q^4)$ term in the Taylor expansions used for approximating
$V(q)$ must be expressed in terms of $\sigma_\mathrm{n}$ or SNR, in
order to be included into the derivation of $d_\mathrm{lim}$.  The
$o(q^4)$ term reflects the error of the Taylor approximation of
$V(q)$. Since $q$ represents a position in the two-dimensional Fourier
plane, the error is proportional to the area $\sigma^2_\mathrm{n}$ of
the region described by $o(q^4)$, modified by the uncertainty
$1/(1-\sigma_\mathrm{n})$ in the border of that region.  Therefore,
$o(q^4)$ can be expressed as follows:
\begin{equation}
o(q^4) = \frac{\sigma^2_\mathrm{n}}{1-\sigma_\mathrm{n}} =
\frac{1}{SNR\,(SNR-1)}\,.
\label{eq:oq4}
\end{equation}
As was discussed above, application of the Taylor expansions is restricted
to the small-size limit, and the representation of $o(q^4)$ given by
Eq.~(\ref{eq:oq4}) can only be used for estimating
$d_\mathrm{lim}$. Estimates of $d_\mathrm{res}$ derived with this
representation would diverge at small SNR.

\section{Resolution limits}
\label{sc:limits}

The resolution criteria given by Eqs.~(\ref{eq:res-cond}) and
(\ref{eq:maxres}) can be applied to derive specific resolution limits
for the templates described above. Analytical solutions can be found
for these equations, following the method outlined in
Sect.~\ref{sc:dlim}--\ref{sc:dres}. If analytical expressions for
$d_\mathrm{lim}$ and $d_\mathrm{res}$ are computationally inconvenient
or cannot be found, the limits can be obtained numerically. Numerical
solutions should then be obtained for an effective $\mathrm{HPBW}=1$,
thus requiring
\begin{equation}
q_\mathrm{max} = \frac{1}{\sqrt{\pi}}\frac{1}{2^{1-\beta/2}}
\end{equation}
to be set for estimating $d_\mathrm{lim}$. For $d_\mathrm{res}$
\begin{equation}
q_\mathrm{res} = \frac{1}{\sqrt{\pi}}\frac{1}{2^{1-\beta/2}} 
\frac{1}{S^{1/(3-\beta)}}
\end{equation}
should be set. Equations~(\ref{eq:res-cond}) and (\ref{eq:maxres}) can
be then solved numerically for $d_\mathrm{lim}$ and $d_\mathrm{res}$
expressed in units of HPBW. The corresponding resolution factors are
obtained from
\begin{equation}
{\cal R}_\mathrm{lim} = d_\mathrm{lim} \frac{\sqrt{\pi}}{2^{2-\beta/2}}\,,
\quad
{\cal R}_\mathrm{res} = d_\mathrm{res} \frac{\sqrt{\pi}}{2^{2-\beta/2}}\,.
\end{equation}
These numerical solutions can be calculated for a relevant range of
SNR, and the dependences of $d_\mathrm{lim}$ and $d_\mathrm{res}$ on
SNR can be established.

This section and Appendix~\ref{sc:app-a} summarize the analytical
resolution limits obtained for the templates described in
Sect.~\ref{sc:templates}.  For each template, general forms of
$d_\mathrm{lim}$ and $d_\mathrm{res}$ and their respective ${\cal
R}_\mathrm{lim}$ and ${\cal R}_\mathrm{res}$ are listed. In all
specific examples, a rectangular, power-law SF given by
Eq.~(\ref{eq:sr}) is adopted, with ${\cal B}_\mathrm{c}$ given by
Eq.~(\ref{eq:hpbw-r}) and the SF correction factor ${\cal
S}_\mathrm{q}$ described by Eq.~(\ref{eq:sq-plaw}). The uniform
weighting ($\beta = 0$) is assumed in all examples. This corresponds,
with a high degree of accuracy, to astronomical images with a Gaussian
PSF.

\subsection{Gaussian component}

\begin{equation}
d_\mathrm{lim} = \frac{2}{\pi}\,
 \left[\ln 2\,
\ln\left(\frac{SNR}{SNR-1}\right)\right]^{1/2} {\cal S}_\mathrm{q}\, {\cal B}_\mathrm{c}\,,
\label{eq:dlim-gauss}
\end{equation}
\begin{equation}
{\cal R}_\mathrm{lim} = \frac{2}{\pi} \left[\ln 2\,
\ln\left(\frac{SNR}{SNR-1}\right)\right]^{1/2}\,.
\label{eq:rlim-gauss}
\end{equation}
For a rectangular, power-law SF, this corresponds to 
\begin{equation}
d_\mathrm{lim} = \frac{2^{2-\beta/2}}{\pi} \left[\pi\,a\,b\ln 2\,
\ln\left(\frac{SNR}{SNR-1}\right)\right]^{1/2}\,.
\label{eq:dlim-gauss-ex}
\end{equation}
For the uniform weighting, Eq.~(\ref{eq:dlim-gauss-ex}) yields a ``benchmark'' value of $d_\mathrm{lim}=1\,\mathrm{HPBW}$ at $\mathrm{SNR}=4$.

\subsubsection{Resolution limit in the image domain}

\begin{figure}
\centerline{
\includegraphics[width=0.45\textwidth, bb =21 35 712 522, clip=true]{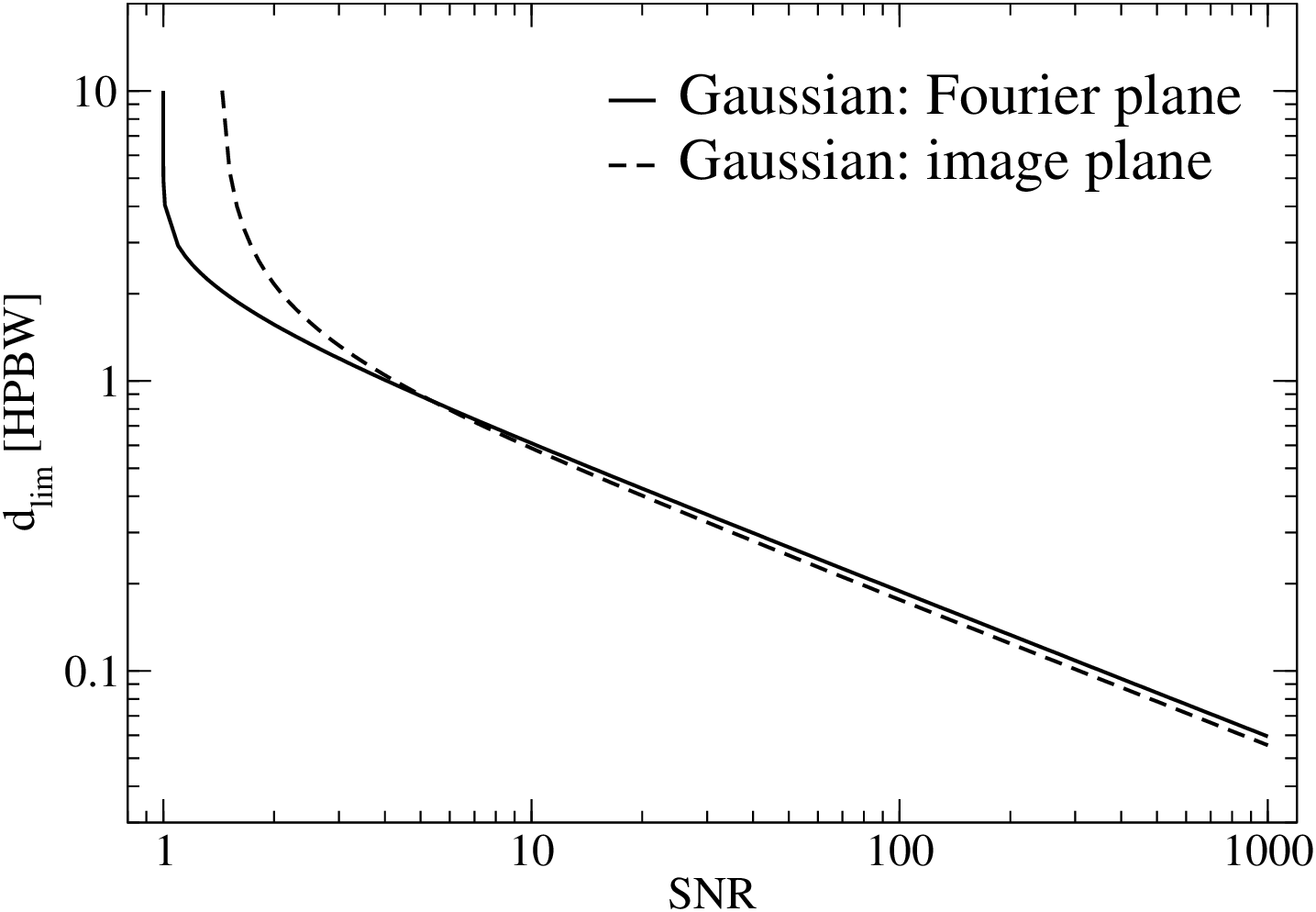}}
\caption{Resolution limits derived for a Gaussian template in the
Fourier plane (solid line) and image plane (dashed line). The
resolution limit in the Fourier domain is derived for a rectangular SF
and the uniform weighting. The resolution limit in the image domain is
derived for a Gaussian PSF. At $\mathrm{SNR}>4$, both limits have the
same dependence $d_\mathrm{lim}\propto SNR^{-1/2}$. At low SNR, the
limit derived in the image plane diverges more rapidly because the
normalization $I_\mathrm{b}(0) = I_\mathrm{f}(0) = 1$ is valid only for
$d_\mathrm{f} \lesssim \mathrm{HPBW}$. }
\label{fg:gauss-fi}
\end{figure}

For the specific case of a Gaussian template fitted to an image with a
Gaussian PSF, $d_\mathrm{lim}$ can be derived in the image plane
as well.  Consider the maximum difference between the Gaussian PSF,
$I_\mathrm{b}(r)$, and a Gaussian template, $I_\mathrm{f}(r)$,
convolved with the PSF. In this case, the deconvolved size,
$d_\mathrm{f}$, of the template can be conveniently expressed in units
of FWHM of the PSF, thus setting the HPBW $d_\mathrm{b} =1$. This
corresponds to a Gaussian with a FWHM $d_\mathrm{i} =
\sqrt{1+d_\mathrm{f}^2}$ fitted directly to the image.  Normalizing
the peak flux density and the PSF to $I_\mathrm{f}(0)= I_\mathrm{b}(0)
= 1$ gives $SNR = 1/\sigma_\mathrm{n}$. This normalization can be
used for all $d_\mathrm{f}$ smaller than, or comparable with, the
HPBW. The PSF is then described by
\begin{equation}
I_\mathrm{b}(r) = C_\sigma \exp \left(-4\ln 2\, \frac{r^2} {1}\right)
\end{equation}
and the template is given by
\begin{equation}
I_\mathrm{f}(r) = C_\sigma \exp \left(-4\ln 2\, \frac{r^2} {1+d_\mathrm{f}^2}\right)\,.
\end{equation}
The template will be resolved in the image if $I_\mathrm{f} - I_\mathrm{b}
\ge \sigma_\mathrm{n}$ at least at one point. The maximum
difference between $I_\mathrm{f}$ and $I_\mathrm{b}$ is realised at
$r=1/\sqrt{2}$, and so the resolution condition becomes
\begin{equation} 
I_\mathrm{f} - I_\mathrm{b} = \left(\frac{1}{4}\right)^{1/(1+d_\mathrm{f}^2)}
- \left(\frac{1}{4}\right) = \frac{\sigma_\mathrm{n}}{C_\sigma} = \frac{1}{C_\sigma\, SNR}\,.
\end{equation}
Solving for $d_\mathrm{lim} = d_\mathrm{f}$ gives the 
\begin{equation}
d_\mathrm{lim} = \left[ \frac{\ln(1/4)}{\ln(\sigma_\mathrm{n}/C_\sigma +1/4)}
- 1\right]^{1/2} = \left[ \frac{\ln(4)}{\ln(\frac{4\,C_\sigma \,SNR}{4 + C_\sigma\,SNR})}
- 1\right]^{1/2}\,,
\label{eq:dlim-gauss-img}
\end{equation}
measured in units of HPBW. This is the deconvolved size of a Gaussian
template, and it can be compared with results from
Eqs.~(\ref{eq:dlim-gauss}--\ref{eq:dlim-gauss-ex}).  For
$\mathrm{SNR}=4$, Eq.~(\ref{eq:dlim-gauss-img}) yields $d_\mathrm{lim}
= 1$\,HPBW, similarly to Eq.~(\ref{eq:dlim-gauss-ex}) applied to
uniformly weighted interferometric data. The two limiting resolutions
are essentially identical at $\mathrm{SNR}>4$ (Fig~\ref{fg:gauss-fi}).
For $\mathrm{SNR}<3$, the asymptotes of Eqs.~(\ref{eq:dlim-gauss})
and (\ref{eq:dlim-gauss-img}) are different, and
Eq.~(\ref{eq:dlim-gauss-img}) diverges at $\mathrm{SNR}=4/3$. This
should be expected, remembering that the normalization used for
$I_\mathrm{b}$ and $I_\mathrm{f}$ can only be used for $d_\mathrm{f}$
(and, consequently, $d_\mathrm{lim}$) smaller than the HPBW (which
necessarily implies $\mathrm{SNR}\ge 4$).

For $d_\mathrm{lim}$ given by Eq.~(\ref{eq:dlim-gauss-img}), the
corresponding limiting size of a Gaussian template convolved with the PSF
is given by
\begin{equation}
d_\mathrm{lim,conv} = \left[ \frac{\ln(4)}{\ln (\frac{4\,SNR}{4+ SNR})}
\right]^{1/2}\,,
\label{eq:dlim-gauss-img-conv}
\end{equation}
which is always larger than the HPBW.

\begin{figure}
\centerline{
\includegraphics[width=0.45\textwidth, bb =31 35 712 522, clip=true]{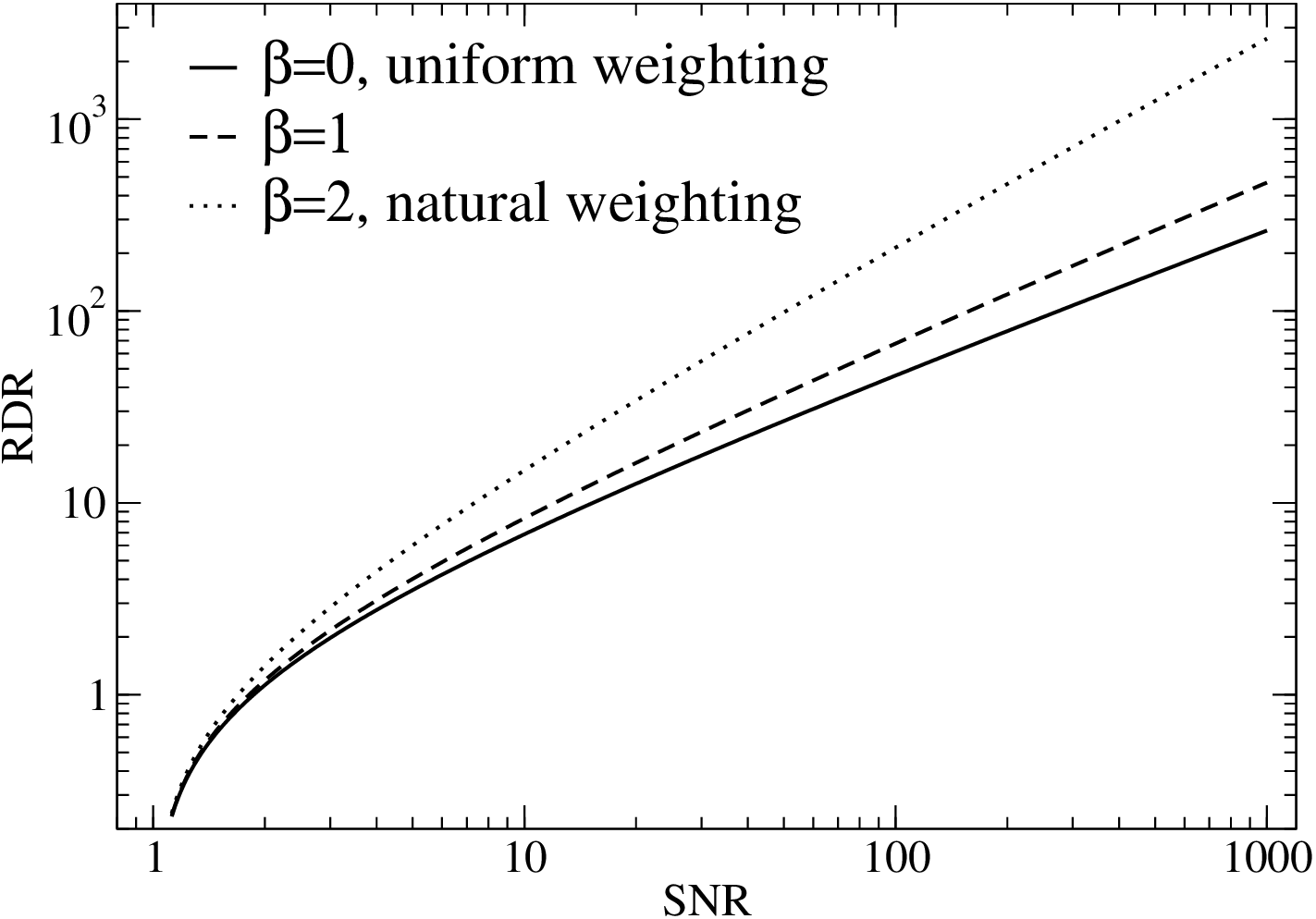}}
\caption{Resolution dynamic range (${\cal D} =
d_\mathrm{res}/d_\mathrm{lim}$) derived for a Gaussian TBD, a
rectangular SF and different values of $\beta$. At high SNR and for poor
Fourier domain coverages, $d_\mathrm{res}$ may exceed $d_\mathrm{lds}$ and should be then substituted by the latter. Valid size estimates are only possible for ${\cal D}\ge 1$. This corresponds to SNR $\ge$ 1.86, 1.82, and 1.70, for $\beta=0,1,2$, respectively.}
\label{fg:resdyn}
\end{figure}

\subsubsection{Maximum resolvable size and the RDR}

\begin{equation}
d_\mathrm{res} = \frac{2}{\pi}\,
 \left[\ln 2\,
SNR^{2/(3-\beta)}\ln\left(SNR\right)\right]^{1/2} {\cal S}_\mathrm{q}\, {\cal B}_\mathrm{c}\,,
\label{eq:dres-gauss}
\end{equation}
\begin{equation}
{\cal R}_\mathrm{res} = \frac{2}{\pi}\left[\ln 2\,
SNR^{2/(3-\beta)}\ln\left(SNR\right)\right]^{1/2}\,.
\label{eq:rres-gauss}
\end{equation}
The maximum resolvable size for a rectangular, power-law SF is given
by
\begin{equation}
d_\mathrm{res} = \frac{2^{2-\beta/2}}{\pi} \left[\pi\,a\,b\, \ln 2\, 
SNR^{2/(3-\beta)}\ln(SNR)\right]^{1/2}\,.
\label{eq:dres-gauss-ex}
\end{equation}
At very high SNR and for poor Fourier domain coverages, $d_\mathrm{res}$ estimated from
Eq.~(\ref{eq:dres-gauss}) can exceed the largest detectable scale
$d_\mathrm{lds} =1/q_\mathrm{min}$. In this case,
$d_\mathrm{lds}$ should be used rather than
$d_\mathrm{max}$. The respective RDR obtained from the ratio of $d_\mathrm{res}/d_\mathrm{lim}$ given by Eqs.~(\ref{eq:dres-gauss}) and (\ref{eq:dlim-gauss-ex}) is then
\begin{equation}
{\cal D} = \left[ SNR^{2/(3-\beta)}\ln(SNR) / \ln\left(\frac{SNR}{SNR-1}\right) \right]^{1/2}\,.
\label{eq:rdr-gauss}
\end{equation}
The RDR is plotted in Fig.~\ref{fg:resdyn} for different values of
$\beta$.  For ${\cal D}\le 1$, no estimates of size can be made.
The SNR levels at which ${\cal D} = 1$ are 1.86, 1.82, and
1.70, for $\beta=0,1,2$, respectively.  At $\mathrm{SNR}\ge 10$, the
RDR is $\propto SNR^{0.76}$, $SNR^{0.84}$, and $SNR^{1.09}$ for
$\beta=0,1,2$, respectively. 

\subsection{Quantum limits on resolution}

\begin{figure}
\centerline{
\includegraphics[width=0.45\textwidth, bb =29 45 714 523, clip=true]{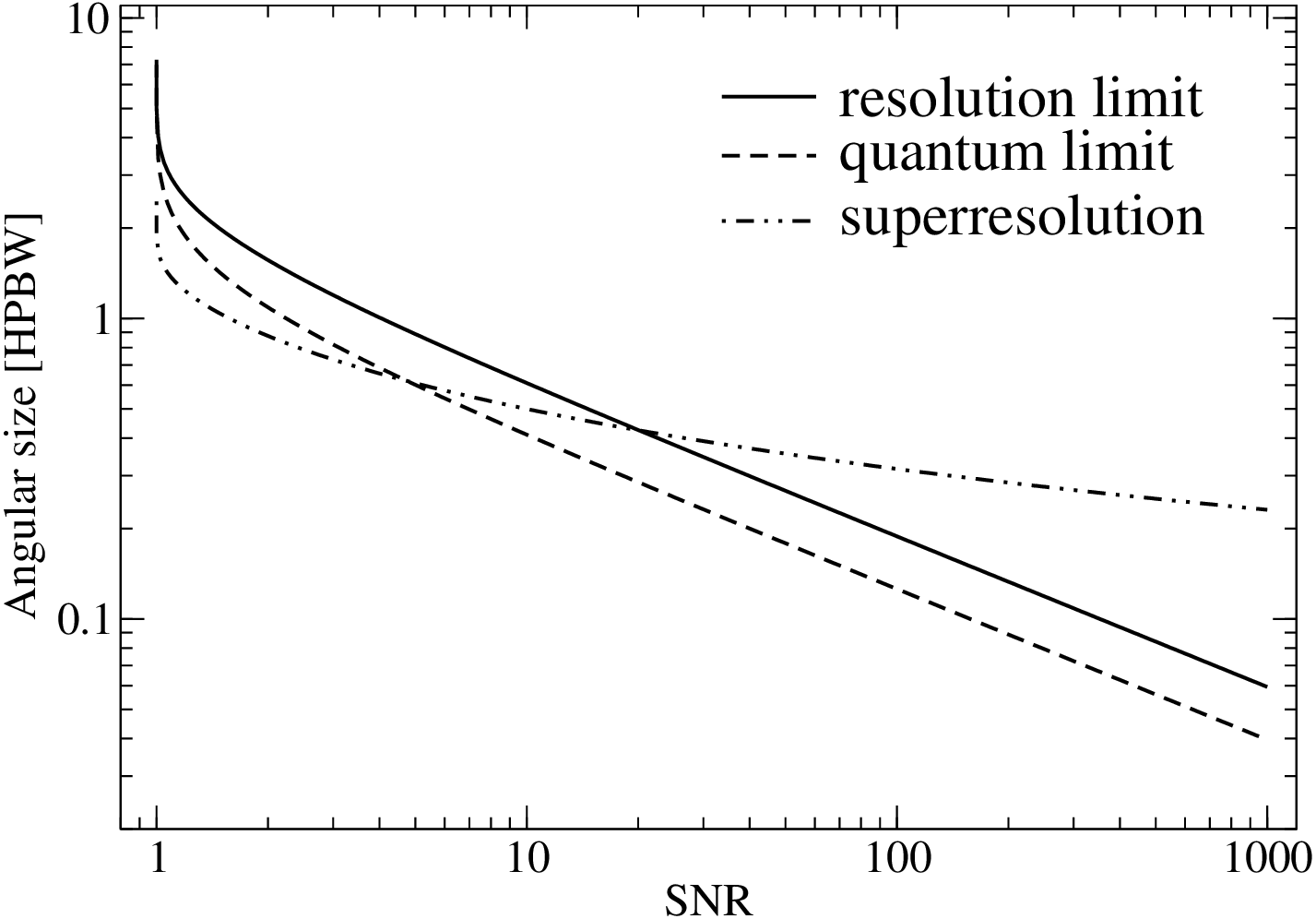}}
\caption{Resolution limit $d_\mathrm{lim}$ for a Gaussian template
compared to the quantum limit $R_\mathrm{q}$ of resolution and the
superresolution limit for out-of-band extrapolation. At
$\mathrm{SNR}>3$, the quantum limit and $d_\mathrm{lim}$ are both
$\propto SNR^{-1/2}$ and $d_\mathrm{lim}\approx 1.5\,R_\mathrm{q}$.  The
superresolution limit approaching $\propto SNR^{-0.1}$ at
$\mathrm{SNR}>10$.  $\mathrm{SNR}=4.76$ marks the lowest limit at which
superresolution can be applied, and $\mathrm{SNR}=20$ gives the
lowest limit for superresolving a Gaussian 
brightness profile.}
\label{fg:qulim}
\end{figure}

The dependence of $d_\mathrm{lim}$ on SNR can be compared with the
ultimate limit on resolution due to quantum fluctuations of light
(Kolobov \& Fabre \cite{kf2000}). The effective quantum resolution
distance $R_\mathrm{q} = R\, S\,/\,(Q+1)$, where $Q+1$ is an effective
number of degrees of freedom (NDF) in the template (Bertero \& De Mol
\cite{bd1996}).  The Rayleigh limit $R = \pi/q_\mathrm{max} =
\pi^{3/2} 2^{1-\beta/2} \sqrt{a\,b}$, and this yields the quantum
limit of resolution
\begin{equation}
R_\mathrm{q} = \pi^{3/2} 2^{1-\beta/2} \frac{S}{Q+1}\,,
\label{eq:qulim}
\end{equation}
expressed in units of HPBW. The values of $S$ and $Q+1$ are calculated
for each SNR and its corresponding $d_\mathrm{lim}$. $R_\mathrm{q}$
represents then the absolute, quantum limit for resolving an emitting
region of finite extent and arbitrary shape detected at a given
SNR. The same equation evaluated at $S = \pi^{-3/2} 2^{(\beta-2)/2}$
gives the SNR dependence of the superresolution limit for out-of-band
extrapolation.  The NDF is evaluated following Kolobov \& Fabre
(\cite{kf2000}) and Frieden (\cite{fri1971}) in the quantum limit and
following Bertero \& De Mol (\cite{bd1996}) in the classical limit
that applies for superresolution.

The quantum and superresolution limits are compared to
$d_\mathrm{lim}$ in Fig.~\ref{fg:qulim}. The quantum limit decreases
at the same rate as $d_\mathrm{lim}$, approaching asymptotically to an 
$SNR^{-1/2}$ proportionality at $\mathrm{SNR}>3$.  This result agrees well
with the theoretical limit on photon number fluctuation in an
interferometer given by $\Delta n\approx n^{-1/2} = SNR^{-1/2}$
(Forward \cite{for1978}, Hariharan \& Sanders \cite{hs1996}).  At
$\mathrm{SNR}>3$, $d_\mathrm{lim}\approx 1.5\,R_\mathrm{q}$, thus
connecting the resolution criterion for a Gaussian template to the
quantum limit.  Compared to the quantum limit and $d_\mathrm{lim}$,
the superresolution limit decreases much slower, approaching an
$SNR^{-0.1}$ dependence at $\mathrm{SNR}>10$. The difference between the
two asymptotes results from different strength of assumptions made
about the extent and shape of the brightness distribution. The
superresolution is not feasible at $\mathrm{SNR} \le 4.76$ for an
object of arbitrary shape. For a Gaussian (or nearly Gaussian)
brightness distribution, application of the superresolution technique
is only feasible at $\mathrm{SNR}\ge 20$.

\subsection{Comparison of resolution limits for different templates}

\begin{figure}
\centerline{ \includegraphics[width=0.45\textwidth, bb =14 35 712 535,
clip=true]{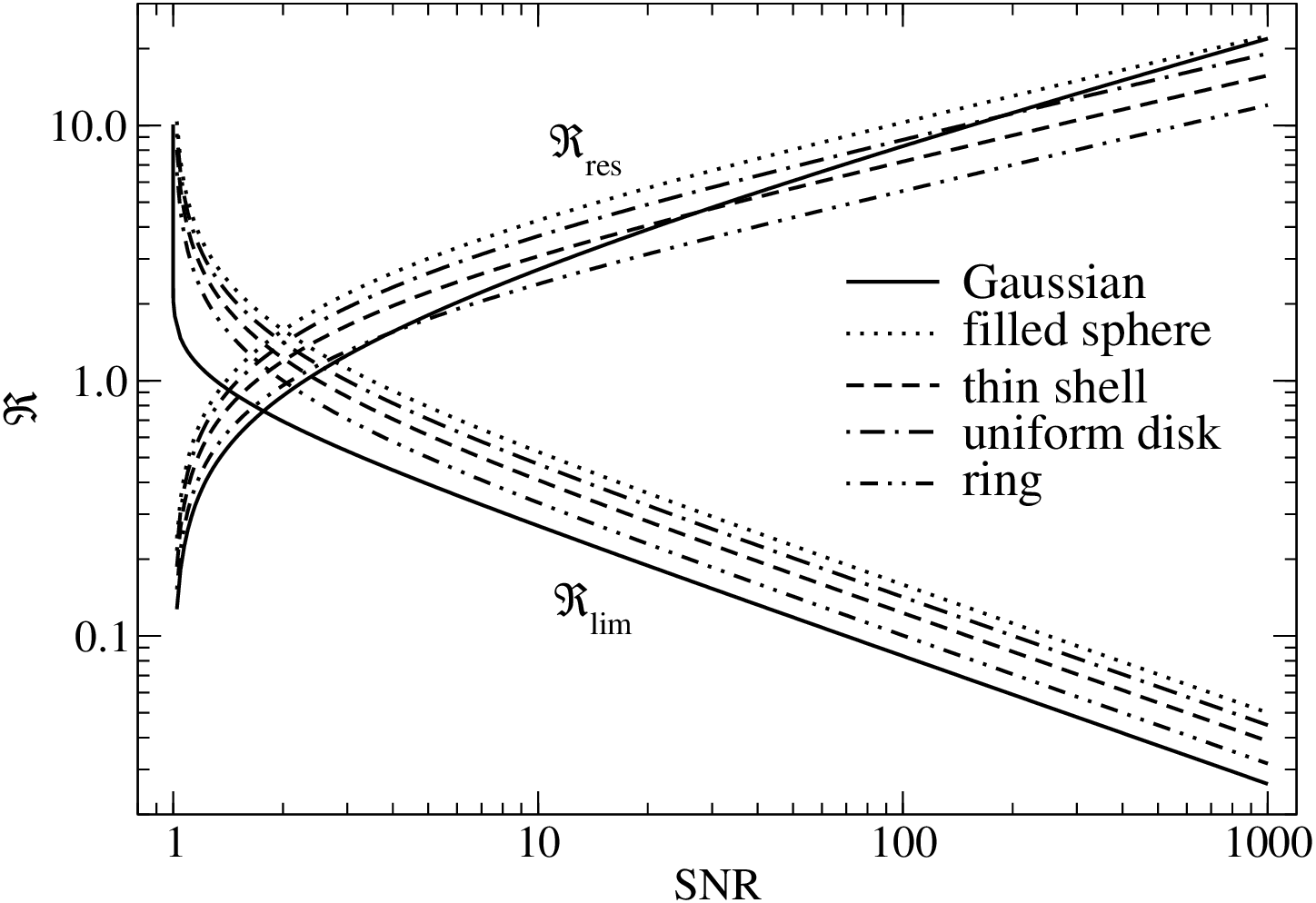}}
\caption{Resolution factors ${\cal R}_\mathrm{lim}$ and ${\cal
R}_\mathrm{res}$ for different templates. At a given SNR, the Gaussian
template has the smallest ${\cal R}_\mathrm{lim}$. All templates have
similar asymptotics ${\cal R}_\mathrm{lim}\propto SNR^{-1/2}$ at
$\mathrm{SNR} \ge 4$. For the largest detectable size and
$\mathrm{SNR}\rightarrow \infty$, the asymptotes are ${\cal
R}_\mathrm{res} \propto SNR^{0.4}$ for a Gaussian template and ${\cal
R}_\mathrm{res} \propto SNR^{0.33}$ for the other templates.  The
steeper rise of ${\cal R}_\mathrm{res}$ of the Gaussian template
compared to the other templates results from the presence of zeros in
the $V(q)$ of the shell, disk, and ring (see Fig.~\ref{fg:vq}).}
\label{fg:rlim}
\end{figure}

Resolution factors and limits for a spherical shell, uniform disk and
ring templates described in Sect.~\ref{sc:templates} are derived in
Appendix~\ref{sc:app-a}.  Resolution factors ${\cal R}_\mathrm{lim}$
and ${\cal R}_\mathrm{res}$ are plotted in Fig.~\ref{fg:rlim} and
compared in Tables~\ref{tb:v-lim} and \ref{tb:v-res} for different
templates and SNR levels. The Gaussian template has the smallest
${\cal R}_\mathrm{lim}$ at all SNR. At $\mathrm{SNR}\ge 4$, ${\cal
R}_\mathrm{lim} \propto SNR^{-1/2}$ for all templates. The picture is
different for ${\cal R}_\mathrm{res}$. At $\mathrm{SNR}\rightarrow
\infty$, the asymptotes are ${\cal R}_\mathrm{res} \propto SNR^{0.4}$
for a Gaussian template and ${\cal R}_\mathrm{res} \propto SNR^{0.33}$
for the other templates. This difference is caused by the presence of
zeros in the $V(q)$ of the shell, disk and ring templates (see
Fig.~\ref{fg:vq}). The Gaussian template reaches zero at infinity, and
thus it can be better detected at large sizes and large SNR, compared
to the other three templates.

\begin{table}[h]
\caption{${\cal R}_\mathrm{lim}$ for different templates}
\label{tb:v-lim}
\begin{center}
\begin{tabular}{l|rrrrrr} \hline\hline
\multicolumn{1}{c|}{${\cal R}_\mathrm{lim}$} & \multicolumn{6}{c}{SNR} \\
         &\multicolumn{1}{c}{1} & \multicolumn{1}{c}{ 3 } & \multicolumn{1}{c}{ 5 } & \multicolumn{1}{c}{ 10 } & \multicolumn{1}{c}{ $10^2$} & \multicolumn{1}{c}{ $10^3$} \\\hline
Gaussian & $\infty$ &  0.337 & 0.250 & 0.172 & 0.053 & 0.016 \\ 
Sphere   & $\infty$ &  0.712 & 0.503 & 0.335 & 0.101 & 0.032 \\ 
Shell    & $\infty$ &  0.551 & 0.390 & 0.260 & 0.078 & 0.024 \\ 
Disk     & $\infty$ &  0.637 & 0.450 & 0.300 & 0.090 & 0.029 \\
Ring     & $\infty$ &  0.450 & 0.318 & 0.212 & 0.064 & 0.020 \\ \hline
\end{tabular}
\end{center}
\end{table}

\begin{table}[h]
\caption{${\cal R}_\mathrm{res}$ for different templates}
\label{tb:v-res}
\begin{center}
\begin{tabular}{l|rrrrrr} \hline\hline
\multicolumn{1}{c|}{${\cal R}_\mathrm{lim}$} & \multicolumn{6}{c}{SNR} \\
         & \multicolumn{1}{c}{ 1 } & \multicolumn{1}{c}{ 3 } & \multicolumn{1}{c}{ 5 } & \multicolumn{1}{c}{ 10 } & \multicolumn{1}{c}{ $10^2$} & \multicolumn{1}{c}{ $10^3$} \\\hline
Gaussian  & 0 & 0.801 & 1.150 & 1.733 &  5.279 & 13.930 \\ 
Sphere   & 0 & 1.396 & 1.915 & 2.701 &  5.540 & 14.171 \\ 
Shell    & 0 & 1.046 & 1.413 & 1.956 &  4.595 &  9.913 \\ 
Disk     & 0 & 1.231 & 1.677 & 2.345 &  5.592 & 12.182 \\
Ring     & 0 & 0.831 & 1.111 & 1.521 &  3.524 &  7.648 \\ \hline
\end{tabular}
\end{center}
\end{table}

\section{Applications}
\label{sc:applications}

\subsection{Brightness temperature estimates from VLBI data}

Brightness temperature, $T_\mathrm{b}$, of non-thermal continuum
emission is commonly used as an indicator of physical conditions in
the emitting material. In relativistic jets, the inverse-Compton limit
of $T_\mathrm{b}\approx 10^{12}$\,K (Kellermann \& Paulini-Toth
\cite{kp1969}, Kellermann \cite{kel2002}) is often reached and
exceeded, implying angular dimensions smaller than the FWHM of the
resolving PSF. In this case, the resolution criteria described by
${\cal R}_\mathrm{lim}$ and $d_\mathrm{lim}$ can be applied to analyze
the size of a template fitted to interferometric visibilities or to
image brightness distribution.  Comparison of $d_\mathrm{lim}$ with
the size obtained from the fit by a template can be used to separate
valid measurements from upper limits.

Elliptical or circular Gaussian components are applied routinely for
estimating sizes of emitting regions. The PSF is often also
represented by an elliptical Gaussian shape. In the most general case
an elliptical Gaussian component is described by its axes
$\theta_\mathrm{maj}$, $\theta_\mathrm{min}$ and the position angle of
the major axis $\psi$. Correspondingly, an elliptical PSF is given by
($b_\mathrm{maj}, b_\mathrm{min}, \phi$). In this case, the resolution
limit should be calculated for $\theta_\mathrm{maj}$ and
$\theta_\mathrm{min}$. The FWHM of the PSF is represented by
its width $b_\psi$ measured along the direction of the axis for which
the limit is calculated. The corresponding resolution limit is then
obtained from Eq.~(\ref{eq:dlim-gauss-ex})
\begin{equation}
\theta_\mathrm{lim,\psi} = 2^{2-\beta/2} b_\psi \left[\frac{\ln 2}{\pi}
\ln\left(\frac{SNR}{SNR-1}\right)\right]^{1/2}\,.
\label{eq:dlim-psi}
\end{equation}
Whenever either of the two axes is smaller than its respective
$\theta_\mathrm{lim,\psi}$, the size obtained from the fit by a
Gaussian component should be treated as upper limit on the size of the
emitting region.

\subsection{Fine structure of young supernovae}

In VLBI observations of early stages of supernova expansion, it is
important to decide at which moment the observed brightness
distribution can be identified with a shell-like structure, thus
distinguishing it from another shape (for instance, a Gaussian or a
uniform disk). Let us assume that the brightness distribution of
emission from a young supernova is fit by a spherical shell of size
$d_\mathrm{s}$ and thickness $\alpha$. The resolution criteria can
then be used to establish whether (1)~the fit is different from a fit
by another template of the same size and (2)~the fit can be distinguished
from a fit by another template of arbitrary size. These two cases form the
necessary and sufficient conditions for verifying the shell-like shape
of the expanding remnant. A method for distinguishing between a
shell and a Gaussian templates is described below. The same procedure can
be used to derive criteria for distinguishing between any other pair of
TBD.

\begin{figure}
\centerline{ \includegraphics[width=0.45\textwidth, bb =13 45 714 522,
clip=true]{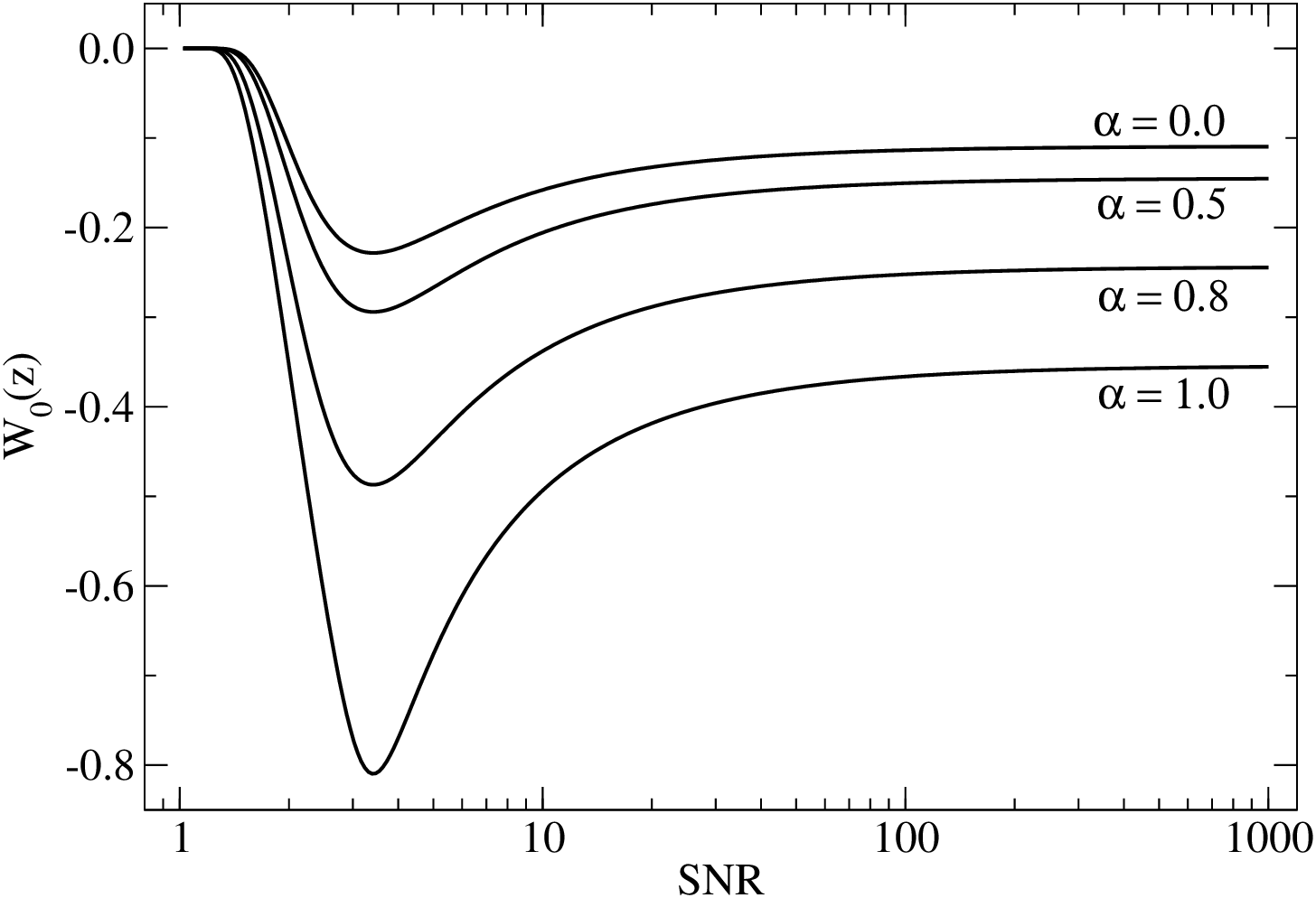}}
\caption{Zero order product logarithm function $W_0(\zeta)$
calculated for a range of shell width parameter $\alpha$.}
\label{fg:lambda0}
\end{figure}

\begin{figure}
\centerline{ \includegraphics[width=0.45\textwidth, bb =31 45 714 522,
clip=true]{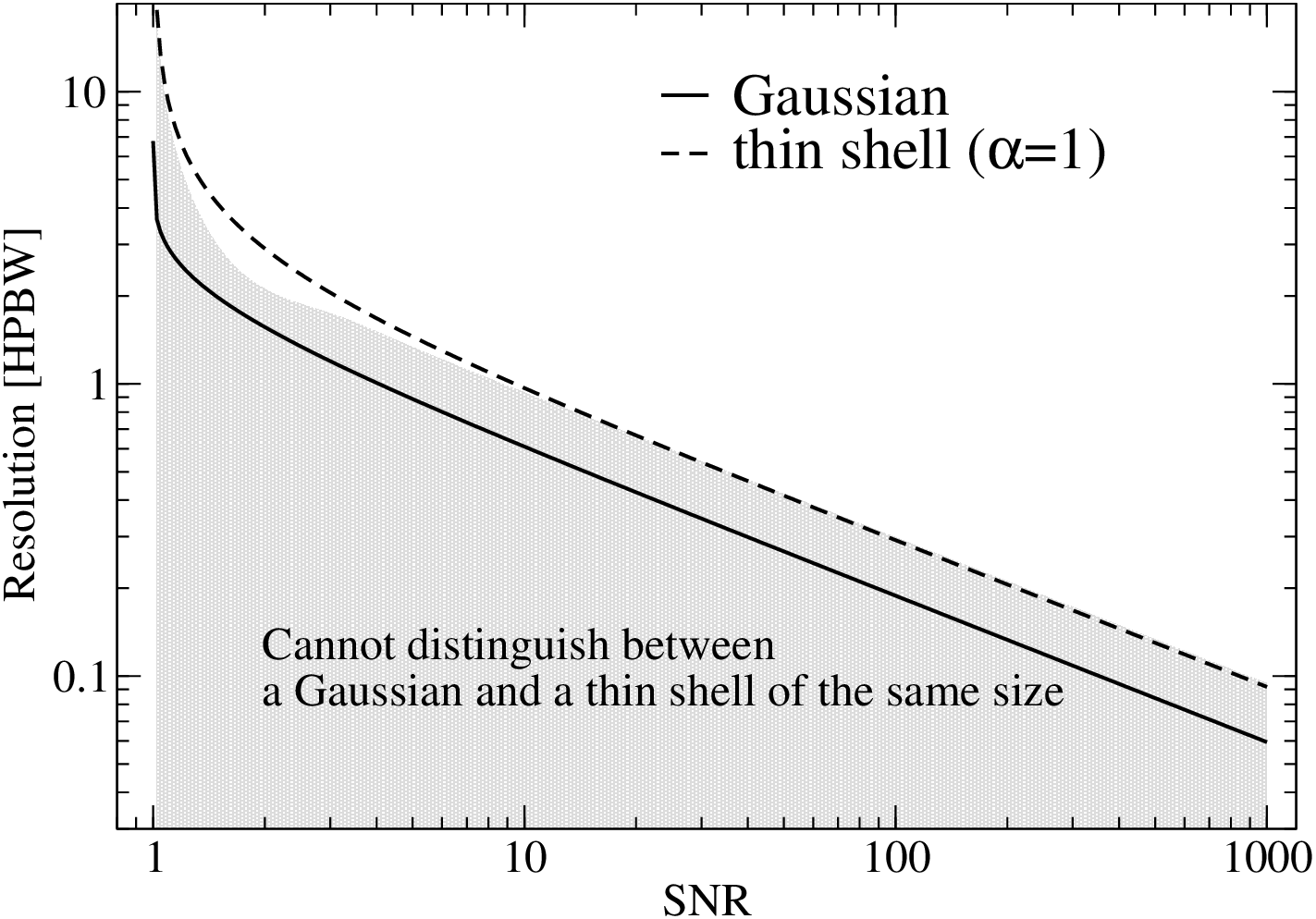}}
\caption{Comparison of the resolution limits for a Gaussian and an
infinitely thin spherical shell. The shaded area marks angular sizes
for which a fit by a thin shell cannot be distinguished from a fit by a
Gaussian of the same size.}
\label{fg:dis-gshell}
\end{figure}

\begin{figure}[h]
\centerline{
\includegraphics[width=0.45\textwidth, bb =31 45 714 523, clip=true]{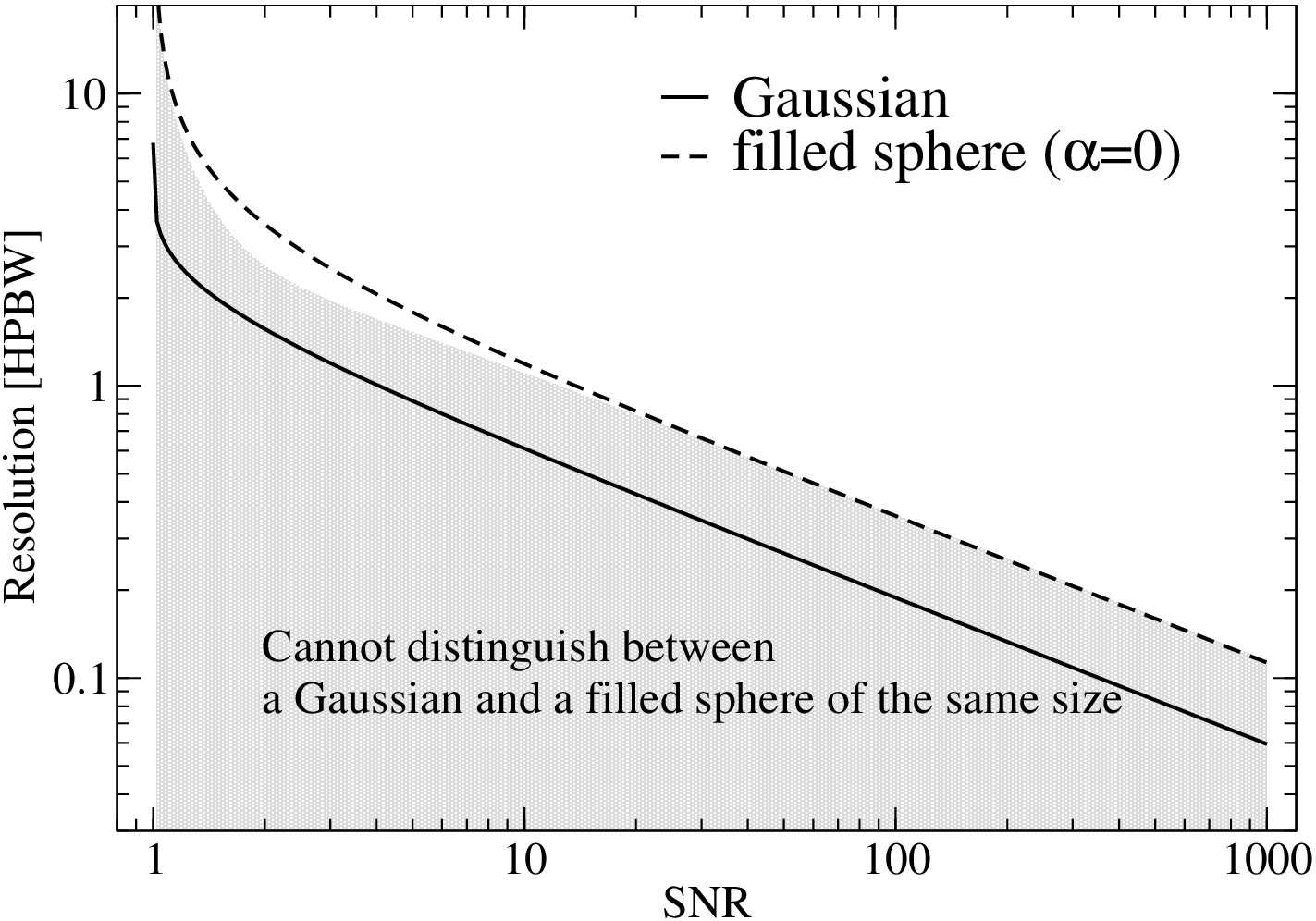}}
\caption{Comparison of the resolution limits for a Gaussian and a
filled sphere. The shaded area marks angular sizes
for which a fit by a filled sphere cannot be distinguished from a fit by a
Gaussian of the same size.}
\label{fg:dis-gsphere}
\end{figure}

To distinguish between a shell and a Gaussian component of the same
size, $d_\mathrm{g,s}$, the condition
\begin{equation}
f(V_\mathrm{s},V_\mathrm{g}) = |\max(V_\mathrm{s} -
V_\mathrm{g})|_{0}^{q_\mathrm{max}} = \sigma_\mathrm{n} = 1/SNR\,
\label{eq:gauss-shell1}
\end{equation}
must be satisfied, where both the Gaussian ($V_\mathrm{g}$) and shell
($V_\mathrm{s}$) visibility amplitudes are normalized to unity at
$q=0$, as given by Eqs.~(\ref{eq:Fu}) and ({\ref{eq:vis-shell1}).
This condition applies only to $d_\mathrm{g,s}$ comparable or smaller
than the HPBW.  Setting $\mathrm{HPBW}\equiv1$, the visibility
difference in Eq.~(\ref{eq:gauss-shell1}) is maximized at
$q_\mathrm{g,s} = 1/\sqrt{4\pi}$.
Equation~(\ref{eq:gauss-shell1}) can then be rewritten as
\begin{equation}
\exp(\frac{1}{\lambda_1} \delta_\mathrm{d}^2) + \frac{1}{\lambda_2} 
\delta_\mathrm{d}^2 = \frac{1}{1-\sigma} = \frac{(SNR-1)^2 + 1}{SNR(SNR-1)}\,,
\label{eq:gauss-shell2}
\end{equation}
where $\delta_\mathrm{d} = \sqrt{d_\mathrm{g,s}^2/(ab)}$ is the relative size of a template measured in units of HPBW, and
$\lambda_{1,2}$ are given by:
\[
\lambda_1 = \frac{16 \ln 2}{\pi}\,,\quad\quad\quad
\lambda_2 = \frac{40}{\pi}\left(\frac{1-\alpha^3}{1-\alpha^5}\right)\,.
\]
Denote 
\[
\xi_\mathrm{SNR} = \frac{(SNR-1)^2 + 1}{SNR(SNR-1)}\,.
\]
Solving Eq.~(\ref{eq:gauss-shell2}) for $\delta_\mathrm{d}$ and recovering
the angular size $\sqrt{\delta_\mathrm{d}^2 a\, b}$ yields the minimum size
at which it is possible to distinguish between the spherical and Gaussian
shapes
\begin{equation}
d_\mathrm{lim,gs} = \sqrt{\frac{a\, b}{SNR}}
\left[\lambda_2 \xi_\mathrm{SNR} +
\lambda_1 {W}_0(\zeta)\right]^{1/2}\,,
\label{eq:dlim-gs}
\end{equation}
where
\[
\zeta = -\frac{\lambda_2}{\lambda_1} 
\exp\left[-\frac{\lambda_2}{\lambda_1} \xi_\mathrm{SNR}\right]
\]
and ${W}_0$ is the zero order product logarithm function
described in Appendix~\ref{sc:app-b}. Figure~\ref{fg:lambda0} shows the dependence of
$W_0(\zeta)$ on SNR and the shell width $\alpha$.
The relevant range of $SNR=[1,\,\infty]$ determines the range of
$\zeta$ $(-1/e,\,0)$. Within this range, an
approximation
\[
W_0(\zeta) + (1 +\frac{1}{e})(\zeta + \frac{1}{e})^{1/e} - 1
\]
can be used.
The resulting limits on resolution are shown in
Figs.~\ref{fg:dis-gshell}--\ref{fg:dis-gsphere} for a thin shell and
filled sphere, respectively. In each figure, the shaded area indicate
sizes for which it is not possible to distinguish between a Gaussian
and a spherical shape of the same size. At large SNR levels this limit
approaches the resolution limit for a spherical shape.

For a Gaussian and a shell of arbitrary sizes $d_\mathrm{g}$,
$d_\mathrm{s}$, the visibility difference reaches a maximum at
$q_\mathrm{gs} \le q_\mathrm{max}$, which is determined by solving the
equation $\mathrm{d}f^2(V_\mathrm{g},V_\mathrm{s})/\mathrm{d}q = 0$.
For each given $d_\mathrm{s}$ (or $d_\mathrm{g}$, reciprocally), the
minimum SNR at which one template can be distinguished from another is
given by $2/f(V_\mathrm{s},V_\mathrm{g})$ evaluated at
$q_\mathrm{gs}$. Below this SNR, it is not possible to claim that the
template fit is unique. Application of this method is exemplified in
Figs.~\ref{fg:dis-gshell-all}--\ref{fg:dis-gsphere-all} for a thin shell
and filled sphere matched against a Gaussian. Similar approach can be
applied to any other pair of templates.

\begin{figure}
\centerline{
\includegraphics[width=0.45\textwidth, bb =31 45 714 522, clip=true]{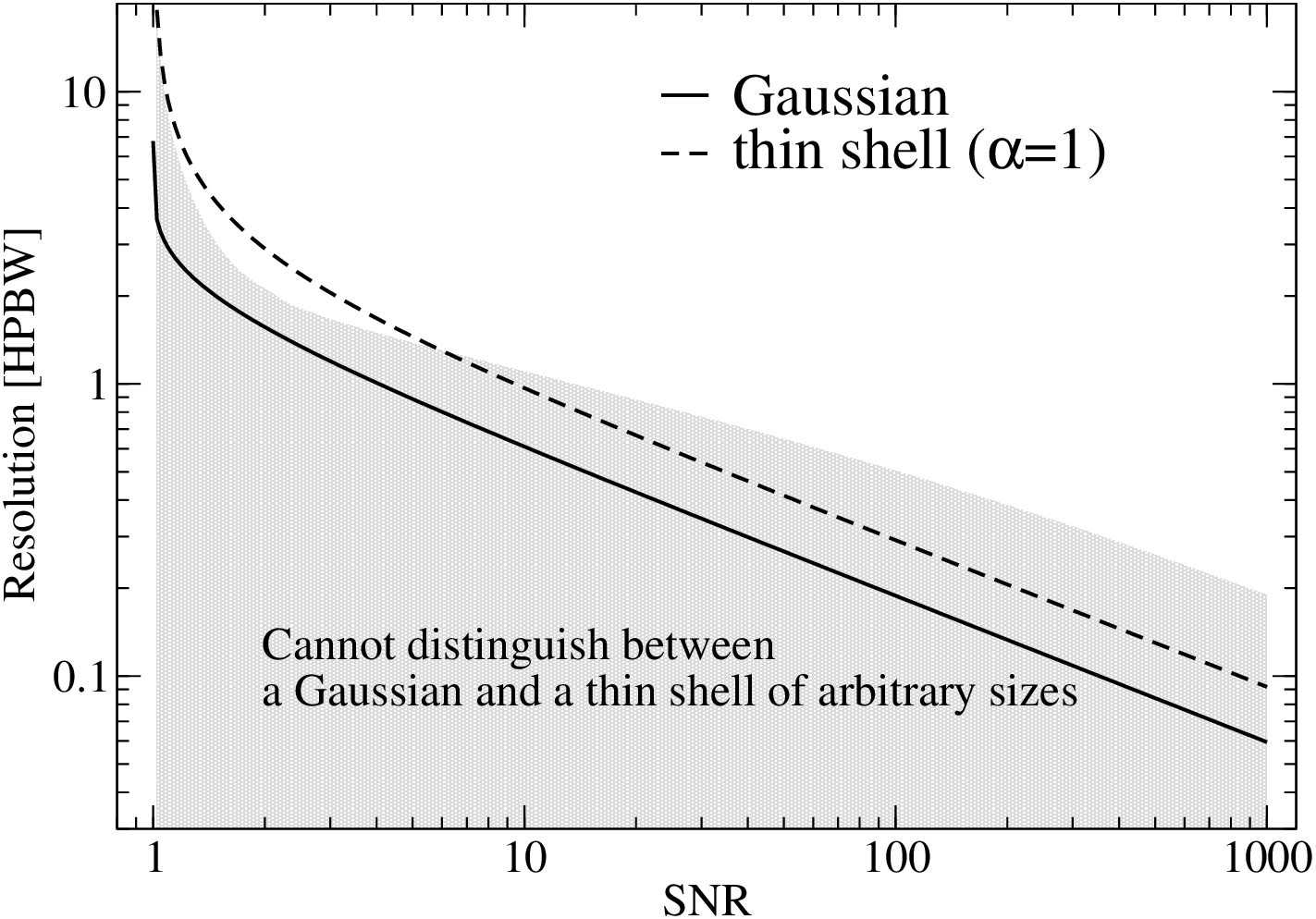}}
\caption{Comparison of the resolution limits for a Gaussian and an
infinitely thin spherical shell. The shaded area marks angular sizes
for which a fit by a thin shell cannot be distinguished from a fit by a
Gaussian of arbitrary size.}
\label{fg:dis-gshell-all}
\end{figure}

\begin{figure}
\centerline{
\includegraphics[width=0.45\textwidth, bb =31 45 714 523, clip=true]{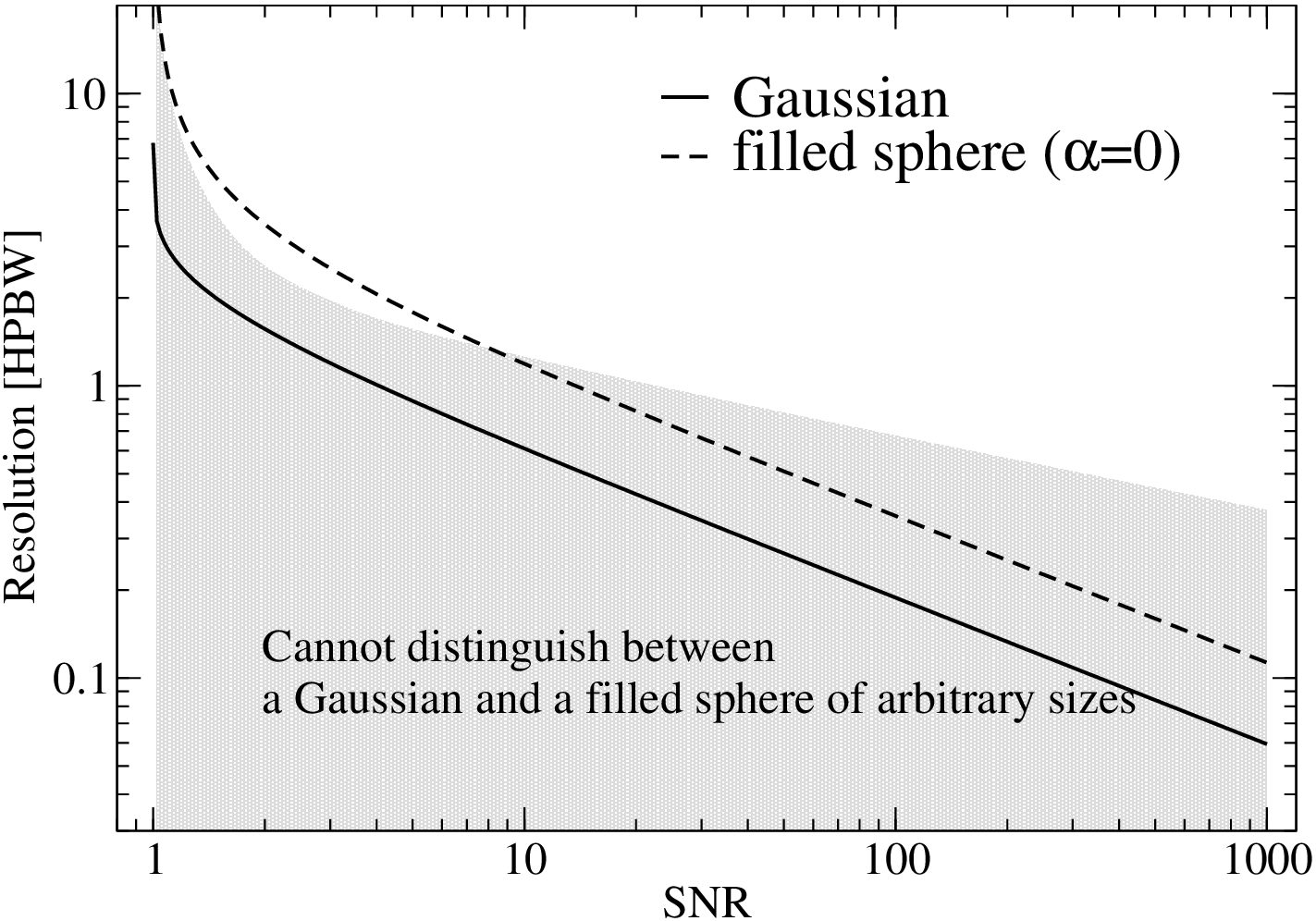}}
\caption{Comparison of the resolution limits for a Gaussian and a
filled sphere. The shaded area marks angular sizes
for which a fit by a filled sphere cannot be distinguished from a fit by a
Gaussian of arbitrary size.}
\label{fg:dis-gsphere-all}
\end{figure}

\appendix

\section{Resolution limits for selected templates}
\label{sc:app-a}

This appendix summarizes analytical expressions derived for the
resolution factors ${\cal R}_\mathrm{lim}$ and ${\cal R}_\mathrm{res}$
and limiting sizes $d_\mathrm{lim}$ and $d_\mathrm{res}$.  The
analytical expressions for ${\cal R}_\mathrm{res}$ and
$d_\mathrm{res}$ are accurate within 40\%. Empirical correction
factors $\kappa(SNR)$ are introduced to
correct this inaccuracy. The correction reduces the errors to within
0.01\% for $\mathrm{SNR}\le 1000$. At larger SNR, the correction
factors become essentially constant.

\subsection{Spherical shell of finite thickness}

\begin{equation}
d_\mathrm{lim} =\frac{2}{\pi}
\left[\frac{1-\alpha^3}{1-\alpha^5} \frac{5}{2(SNR-1)}\right]^{1/2}{\cal S}_\mathrm{q} \, {\cal B}_\mathrm{c}\,,
\label{eq:dlim-sphere}
\end{equation}
\begin{equation}
{\cal R}_\mathrm{lim} = \frac{2}{\pi}\left[\frac{1-\alpha^3}{1-\alpha^5} \frac{5}{2(SNR-1)}\right]^{1/2}\,.
\label{eq:rlim-sphere}
\end{equation}
For a rectangular, power-law SF and uniform weighting, the corresponding limiting resolution for a spherical shell becomes
\begin{equation}
d_\mathrm{lim} = \sqrt{\frac{40}{\pi}  \left(\frac{1 - \alpha^3}{1-\alpha^5}
\right) \frac{a\,b}{SNR-1}}\,.
\label{eq:dlim-sphere-ex}
\end{equation}
The two limiting cases of the spherical shell are described by the filled
sphere ($\alpha=0$) and infinitely thin shell ($\alpha\rightarrow 1$),
which gives
\begin{equation}
d_\mathrm{lim,sphere}  = \sqrt{\frac{40}{\pi} 
\frac{a\,b}{SNR-1}}\,,
\label{eq:dlim-ball}
\end{equation}
\begin{equation}
d_\mathrm{lim,shell} =  \sqrt{\frac{24}{\pi}
\frac{a\,b}{SNR-1}}\,.
\label{eq:dlim-thinshell}
\end{equation}
The largest detectable size is given by
\begin{equation}
d_\mathrm{res} = \frac{2}{\pi} \,\Phi(SNR)\,\left[\frac{5}{2} \frac{1-\alpha^3}{1-\alpha^5}
\right]^{1/2}\,{\cal S}_\mathrm{q} \, {\cal B}_\mathrm{c}\,,
\end{equation}
\begin{equation}
{\cal R}_\mathrm{res} =  \frac{2}{\pi}\,\Phi(SNR)\,\left[\frac{5}{2} \frac{1-\alpha^3}{1-\alpha^5}
\right]^{1/2}\,,
\end{equation}
where
\[
\Phi(SNR) = SNR^{1/(3-\beta)}\left(\frac{SNR-1}{SNR}\right)^{1/2}\,.
\]

\subsection{Disk of uniform brightness}

The resolution limit is given by
\begin{equation}
d_\mathrm{lim} = \frac{2}{\pi}
\left[ \frac{2}{SNR-1}\right]^{1/2}{\cal S}_\mathrm{q} \, {\cal B}_\mathrm{c}
\,,
\end{equation}
\begin{equation}
{\cal R}_\mathrm{lim} = \frac{2}{\pi}\left[ \frac{2}{SNR-1}\right]^{1/2}\,.
\end{equation}
The largest detectable size is determined by
\begin{equation}
d_\mathrm{res} = \frac{2}{\pi}\, \sqrt{2}\,\Phi(SNR)\,{\cal S}_\mathrm{q} \, {\cal B}_\mathrm{c}\,,
\end{equation}
\begin{equation}
{\cal R}_\mathrm{res} =  \frac{2}{\pi}\,\sqrt{2}\,\Phi(SNR)\,.
\end{equation}

\subsection{Ring}

The resolution limit is given by
\begin{equation}
d_\mathrm{lim} = \frac{2}{\pi}
\left[\frac{1}{SNR-1}\right]^{1/2}{\cal S}_\mathrm{q} \, {\cal B}_\mathrm{c}\,,
\end{equation}
\begin{equation}
{\cal R}_\mathrm{lim} = \frac{2}{\pi}\left[\frac{1}{SNR-1}\right]^{1/2}\,.
\end{equation}
The largest detectable size is described by
\begin{equation}
d_\mathrm{res} = \frac{2}{\pi}
\Phi(SNR)\,{\cal S}_\mathrm{q} \, {\cal B}_\mathrm{c}\,,
\end{equation}
\begin{equation}
{\cal R}_\mathrm{res} = \frac{2}{\pi}\Phi(SNR)\,.
\end{equation}

\subsection{Empirical corrections for $d_\mathrm{res}$ and ${\cal R}_\mathrm{res}$}

Analytical expressions for $d_\mathrm{res}$ and ${\cal
R}_\mathrm{res}$ must be corrected for the error of the Taylor expansion
in the large-size limit. Thus, for the shell, disk, and ring templates
\begin{equation}
{\cal R}_\mathrm{res,corr} = \kappa(SNR)\,{\cal R}_\mathrm{res}\,,
\end{equation} 
with the empirical correction factors $\kappa(SNR)$ given by 
\begin{equation}
\begin{split}
\kappa(SNR)& = 1 + a_1\,\lg\left(a_2\, \lg(SNR)+ 1\right) + \\
  & + a_3 \lg\left(
\frac{1}{a_4 [\lg(SNR)]^2 + 1}\right)\,. \\
\end{split}
\end{equation}
The coefficients $a_{1}$--$a_{4}$ are listed in Table~\ref{tb:coef}.
At $\mathrm{SNR}\ge 300$, $\kappa(SNR)\approx const$, for all templates.
The limiting values of $\kappa(SNR)$ are given in Table~\ref{tb:coef}.
\begin{table}[h]
\caption{Coefficients for $\kappa(SNR)$}
\label{tb:coef}
\begin{center}
\begin{tabular}{c|cccc} \hline\hline
 & Sphere & Shell & Disk & Ring \\\hline
$a_1$ & $8.95923$ & $6.30926$ & $6.87495$ & $3.24248$ \\
$a_2$ & $0.11622$ & $0.13358$ & $0.13955$ & $0.20988$  \\
$a_3$ & $2.23652$ & $1.34282$ & $1.69345$ & $0.82607$  \\
$a_4$ & $0.12907$ & $0.22302$ & $0.17399$ & $0.31843$  \\
$\kappa(SNR>300)$ & 1.42    &  1.28    &  1.35   & 1.20    \\\hline
\end{tabular}
\end{center}
\end{table}

\section{Product logarithm function}
\label{sc:app-b}

The zero order product logarithm function ${W}_0$
(Fig.~\ref{fg:prodlog}) provides the principal solution for $w$ in the
equation $w \exp(w) = z(w)$.  $W_0$ is the principal value of the
Lambert $W$-function (Corless et al. \cite{cor+1996}), and it is
essentially an extension of the logarithm function. $W_0$ is
real-valued for $z>-1/e$.

\begin{figure}
\centerline{ \includegraphics[width=0.45\textwidth, bb =16 51 706 522,
clip=true]{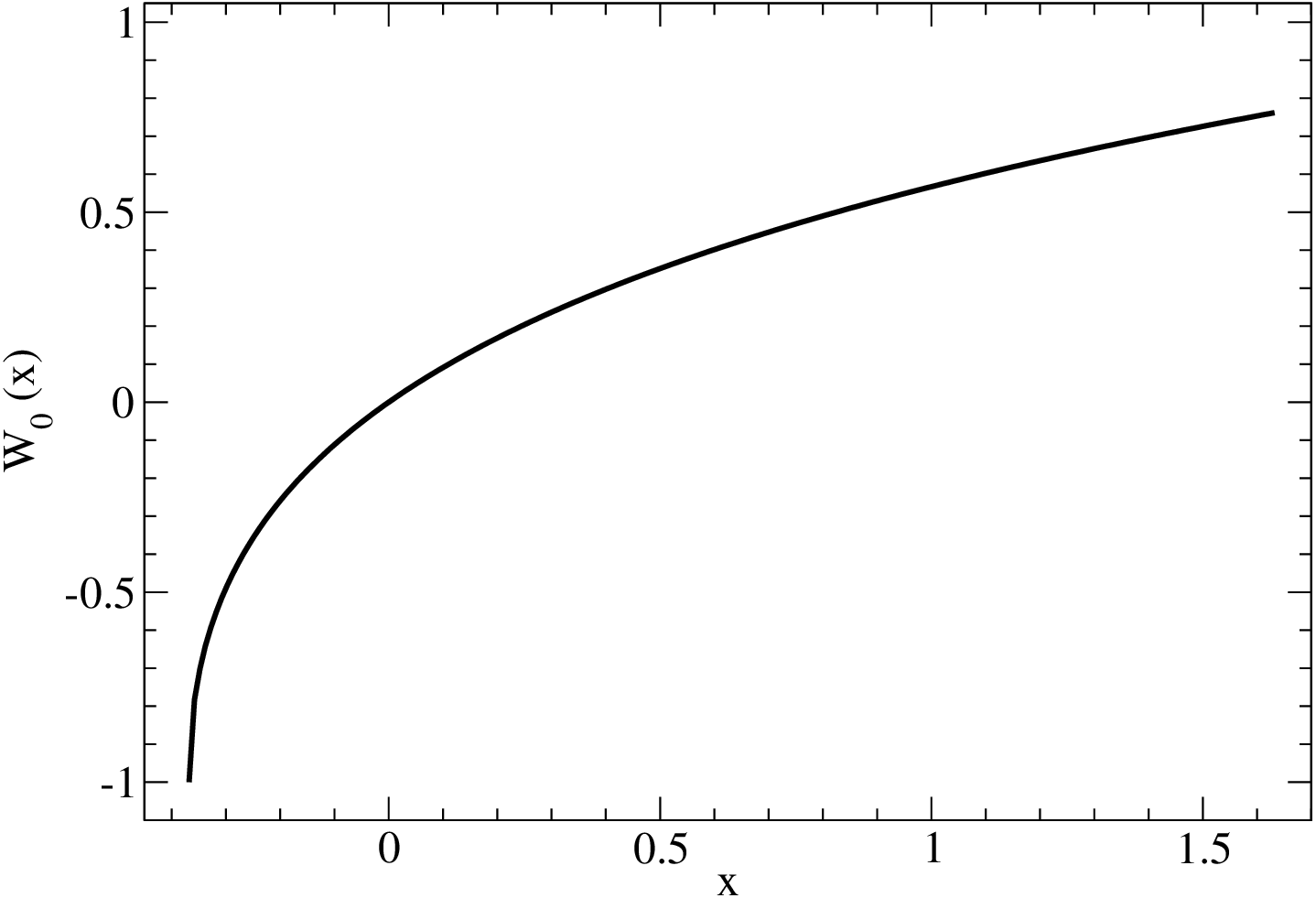}}
\caption{Zero order product logarithm function $W_0$ plotted for
a range of arguments that cover the small and the large argument
asymptotics.}
\label{fg:prodlog}
\end{figure}


\begin{thebibliography}{99}

\bibitem[2003]{bb2003} Bartel, N., Bietenholz, M.~F. 2003, ApJ, 591,
301

\bibitem[1996]{bd1996} Bertero, M., De Mol, C. 1996, in Progress in
Optics, ed. E. Wolf (Elsevier: Amsterdam),
v.\,XXXVI, 129

\bibitem[1996]{cor+1996}Corless, R.~M., Gonnet, G.~H., Hare, D.~E.~G.,
Jeffrey, D.~J., Knuth, D.~E. 1996, Adv. Comput. Math. 5, 329

\bibitem[1995]{bri95} Briggs, D.~S. 1995, PhD Dissertation, NMIMT, Socorro NM, USA

\bibitem[1999]{bss99} Briggs, D.~S., Schwab, F.~R., Sramek,
R.~A. 1999, in ASP Conf. Ser., v.\,180, Synthesis imaging in radio
astronomy II, eds. Taylor, G.~B., Carilli, C.~L., Perley, R.~A. (ASP:
San Francisco), 127

\bibitem[1999]{fom91} Fomalont, E.~B. 1999, in ASP Conf. Ser.,
v.\,180, Synthesis imaging in radio astronomy II, eds. Taylor, G.~B.,
Carilli, C.~L., Perley, R.~A. (ASP: San Francisco), 301

\bibitem[1978]{for1978} Forward, R.~L. 1978, Phys. Rev. D, 17, 379

\bibitem[1971]{fri1971} Frieden, B.~R. 1971, in Progress in
Optics, ed. E. Wolf (Elsevier: Amsterdam),
v.\,IX, 311

\bibitem[1996]{hs1996} Hariharan, R., Sanders, B.~C. 1996, in Progress in
Optics, ed. E. Wolf (Elsevier: Amsterdam),
v.\,XXXVI, 49

\bibitem[2004]{hor+2004} Horiuchi, S., Fomalont, E.~B., Taylor, W.~K.,
et al. 2004, ApJ, 616, 110

\bibitem[2002]{kel2002} Kellermann, K.~I. 2002, PASA, 19, 77

\bibitem[1969]{kp1969} Kellermann, K.~I., Paulini-Toth, I.~I.~K. 1969,
ApJ, 155, L71

\bibitem[2000]{kf2000} Kolobov, M.~I., Fabre, C. 2000, Phys. Rev. Lett., 85, 18, 3789
  
\bibitem[2003]{lob03} Lobanov, A.~P. 2003, SKA Memo No.\,38

\bibitem[2000]{lob+2000} Lobanov, A.~P., Krichbaum, T.~P., Graham,
D.~A., et al. 2000, A\&A, 364, 391

\bibitem[1995]{mar+95} Marcaide, J.~M., Alberdi, A., Ros, E., et al. 1995,
Science, 270, 1475

\bibitem[2001]{mcd+2001} McDonald, A.~R., Muxlow, T.~W.~B., Pedlar,
A., et al. 2001, MNRAS, 322, 100

\bibitem[1996]{mol+1996} Moellenbrock, G.~A., Fujisawa, K., Preston,
R.~A., et al.  1996, AJ, 111, 2174

\bibitem[1976]{nb76} Norton, R.~H., Beer, R. 11976, J. Opt. Soc. Amer. 66, 259

\bibitem[1999]{pea99} Pearson, T.~J. 1999, in ASP Conf. Ser., v.\,180,
Synthesis imaging in radio astronomy II, eds. Taylor, G.~B., Carilli,
C.~L., Perley, R.~A. (ASP: San Francisco), 335

\bibitem[2002]{per+2002} P\'erez-Torres, M.~A., Alberdi, A., Marcaide,
J.~M., et al.  2002, MNRAS, 335, L23

\bibitem[1879]{ray1879} Rayleigh, J.~W.~S. 1879, Phil. Mag., 8, 261

\bibitem[2002]{sto2002} Stoughton, C., Lupton, R.~H., Bernardi, M., et
al. 2002, AJ, 123, 485

\bibitem[2002]{td2002} Tango, W.~J., Davis, J. 2002, MNRAS, 333, 642

\bibitem[1999]{tcp99} Taylor, G.~B., Carilli, C.~L., Perley,
R.~A. (eds.), Synthesis imaging in radio astronomy II, ASP
Conf. Ser., v.\,180 (ASP: San Francisco) 

\bibitem[1986]{tms86} Thompson, A.~R., Moran, J.~M., Swensson,
G.~W. 1986, Interferometry and Synthesis in Radio Astronomy, New York:
Wiley.

\bibitem[1969]{tor1969} Toraldo di Francia, G. 1969, J. Opt. Soc. Am., 59, 799
 
\end{thebibliography}
\end{document}